\documentclass[a4,prd, floatfix, 
nofootinbib,
preprintnumbers
,twocolumn
]{revtex4}
\usepackage[utf8x]{inputenc}
\usepackage[T1]{fontenc}
\usepackage{lmodern}
\usepackage{datetime}
\usepackage{amsmath}
\usepackage{amssymb}
\usepackage{bm}
\usepackage{epsfig}
\usepackage{graphicx}
\usepackage{color}
\usepackage{hyperref}
\usepackage{slashed}
\usepackage{empheq}

\newcommand{\be}{\begin{equation}}
\newcommand{\ee}{\end{equation}}
\newcommand{\bdm}{\begin{displaymath}}
\newcommand{\edm}{\end{displaymath}}
\newcommand{\bea}{\begin{eqnarray}}
\newcommand{\eea}{\end{eqnarray}}
\newcommand{\nn}{\nonumber}

\newcommand{\ba}{\begin{align}}
\newcommand{\ea}{\end{align}}

\newcommand{\ipnp}{Institute of Particle and Nuclear Physics,
  Faculty of Mathematics and Physics,
  Charles University in Prague, V Hole\v{s}ovi\v{c}k\'ach 2,
  180 00 Praha 8, Czech Republic}

\begin{document}
\title{Universal global analytic expansion for the 't~Hooft-Polyakov monopole profiles}
\preprint{}
\pacs{\bf PACS should be added at some point}
\author{Michal Malinsk\'{y} 
}
\email{michal.malinsky@matfyz.cuni.cz}
\affiliation{\ipnp}
\begin{abstract}
In this work we discuss in detail a global analytic expansion scheme for the solutions of the `t~Hooft-Polyakov monopole profile equations for arbitrary $\lambda/e^2>0$ based on the findings presented in a recent  resurgence-oriented letter~{\tt arXiv:2602.14620 [hep-th]}, which the present study significantly expands upon. A uniformly convergent functional perturbation series developed around universal, surprisingly simple, analytic non-perturbative background profiles corresponding to a partial resummation of the Borel-plane expansions suggested there, is constructed; a perfect match to what is known about the full solutions' local behaviour at zero and infinite radii is achieved, along with simple analytic prescriptions for the locally inaccessible numerical parameters therein.           
\end{abstract}
\maketitle
\section{Introduction\label{sec:introduction}}
Magnetic monopoles, topologically non-trivial extended gauge and scalar field configurations occurring naturally in beyond-Standard model theories with simple or semi-simple gauge groups (such as Grand unifications~\cite{Georgi:1974sy} or some of their direct descendants), are among very few direct probes to their global properties that perturbative baryon and lepton number violating processes such as proton decay sample only locally.
The canonical setting in which these objects are traditionally studied is the so called 't~Hooft-Polyakov monopole~\cite{tHooft:1974kcl,Polyakov:1974ek} corresponding to a gauge model with an $SO(3)\to SO(2)$ symmetry breaking (with desired topological properties such as non-triviality of its second homotopy group, $\pi_2(SO(3)/SO(2))=\mathbb{Z}$, cf.~\cite{PhysRevD.50.2806}) triggered by a real scalar triplet field. 

Their intriguing radially symmetric spatial profiles are known to obey a pair of second-order ordinary differential equations (ODE's)
\begin{eqnarray}\label{eq:gaugeprofile}
y''&=&y \,z^2 +y(y^2-1)/x^2\,,\\
\label{eq:scalarprofile}
z''+2z'/x &=&2 z \,y^2/x^2+\beta z(z^2-1)\,,
\end{eqnarray}
where $\beta\equiv \lambda/e^2$ is a combination of the only two physical couplings in the problem, namely, the scalar-sector quartic self-coupling $\lambda$ and the gauge coupling $e$; the only scale present, corresponding to the $C$ factor of Ref.~\cite{Prasad:1975kr}, has been conveniently set to 1.
Note also that the independent variable $x$ and the two profile functions $y$ and $z$ in Eqs.~(\ref{eq:gaugeprofile})-(\ref{eq:scalarprofile}) correspond, respectively, to  $x\leftrightarrow r$, $y(x)\leftrightarrow K(r)$ and $x z(x)\leftrightarrow H(r)$ in the original notation therein. The work~\cite{Prasad:1975kr} is also where the infamous Bogomolny-Prasad-Sommerfield (BPS) analytic solution 
\be\label{eq:BPS}
y_{\rm BPS}(x)={x}/{\sinh x}\,,\quad z_{\rm BPS}(x)=\coth x-{1}/{x}\,,
\ee
obeying the physical boundary conditions 
\bea
\label{eq:boundaryconditionsy}
y({x\to 0})\to 1\,,\; 
y({x\to \infty}) \to  0\,, \\
\label{eq:boundaryconditionsz}
z({x\to 0})\to 0\,,\; 
z({x\to \infty}) \to  1\,, 
\eea
in the ``ideal'' (albeit slightly unphysical) $\beta = 0$ setting
had been first found by the method of ``shimmying'', see also~\cite{Bogomolny:1975de}. 
Note that the conditions~(\ref{eq:boundaryconditionsy})-(\ref{eq:boundaryconditionsz}), dictated by the non-trivial topology and finite-energy requirements, are generally rather difficult to fulfil due to the unpleasant singularities both Eqs.~(\ref{eq:gaugeprofile})-(\ref{eq:scalarprofile}) feature at the origin.

\begin{figure}[t]
  \includegraphics[width=0.42\textwidth]{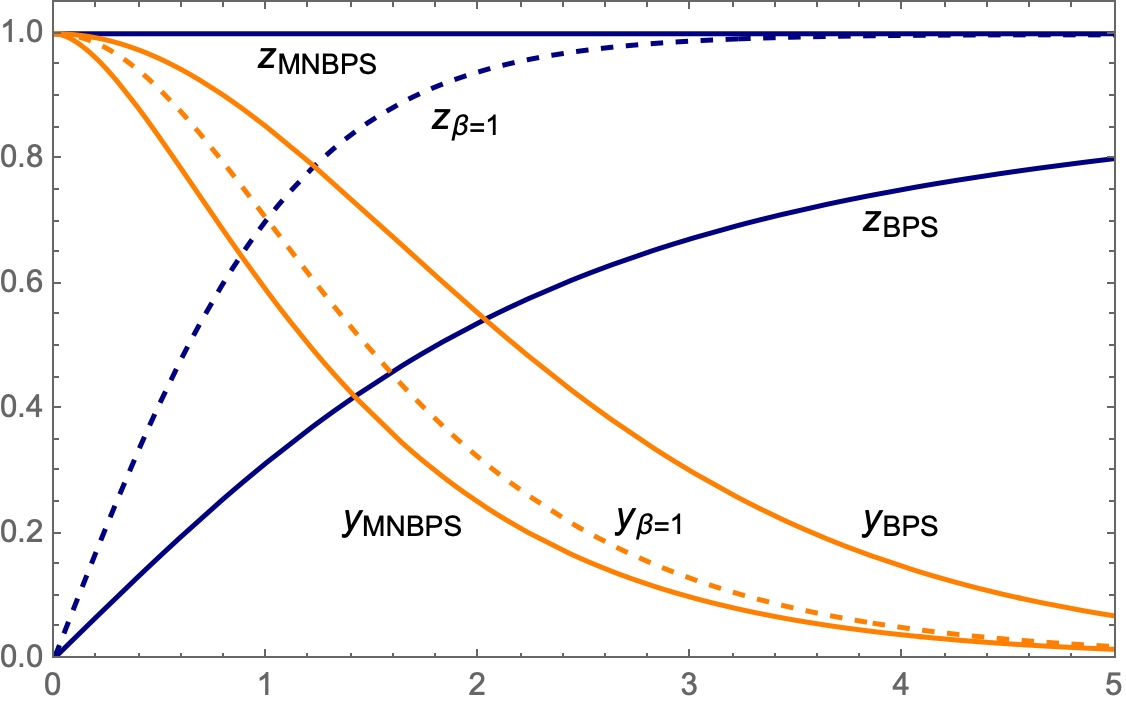}\;
\caption{\label{fig:numericsinintro}
Gauge (in orange) and scalar (in blue) radial profiles of three sample 't~Hooft-Polyakov monopole settings corresponding to $\beta=0$ (BPS) and $\beta\to \infty$ (MNBPS) in solid lines, and $\beta=1$ (specific NBPS) in dashed lines; all these are governed by ODE's~(\ref{eq:gaugeprofile})-(\ref{eq:scalarprofile}) and boundary conditions~(\ref{eq:boundaryconditionsy})-(\ref{eq:boundaryconditionsz}).} 
\end{figure}

Concerning the general non-BPS (NBPS) $\beta>0$ cases, no closed form solutions are known, not even in the structurally simplified $\beta\to \infty$ limit (called maximally non-BPS (MNBPS) in Ref.~\cite{Malinsky:2026eux}) in which the scalar profile becomes trivial, $z(x)=1$ $\forall x>0$, and Eq.~(\ref{eq:gaugeprofile}) reduces to
\begin{equation}\label{eq:MNBPSprofile}
y''=y+(y^3-y)/x^2\,.
\end{equation}
Hence, most of the information we currently have about the corresponding solutions comes from various types of semi-analytical or purely numerical accounts based, for instance, on controlled small-parameter/local expansions, or even vanilla ODE integrators (see, for instance, \cite{Julia:1975,Bais:1976,Goddard:1977da,Gardner:1983,Breitenlohner:1992,Forgacs:2005vx} and references therein); for illustration, a set of sample numerical shapes of $y$ and $z$ in the $\beta=0$, $\beta=1$ and $\beta\to\infty$ settings is depicted in Fig.~\ref{fig:numericsinintro}.

As for purely analytical accounts of NBPS settings, one is generally limited by the the radius of convergence of different types of local expansions at finite $x$, which even degenerates to zero for the asymptotic expansions at $x\to \infty$ that the system~(\ref{eq:gaugeprofile})-(\ref{eq:scalarprofile}) admits. As for the former, perhaps the most interesting of these is the power-log transseries expansion of the MNBPS profile at $x=0$ reading (in the notation of Ref.~\cite{Malinsky:2026eux})
\begin{equation}\label{eq:Frobeniusat0intro}
y(x)=\sum_{m=0}^\infty \sum_{n=0}^m b_{m,n}x^{2m}(\log x)^n\,,
\end{equation}
where $b_{0,0}=1$, $b_{1,0}=B_\infty$, $b_{1,1}=\tfrac{1}{3}$ etc., with $B_\infty$ denoting a real parameter that can not be determined 
locally. Interestingly, Frobenius logs emerge there due to a resonance between the linear part of the ODE~(\ref{eq:MNBPSprofile})  and the ``overly simple'' $y^3/x^2$ non-linearity therein; the finite-$\beta$ settings generally do not feature these, cf.~\cite{Dunne:2026hfx}. 
Similarly, the asymptotic expansion of the MNBPS profile around $x\to \infty$ is also known,
\begin{equation}\label{eq:InfinityTransseriesintro}
y(x)= \sum_{m=0}^\infty \sum_{n=0}^m  a_{m,n}x^{-m}{e^{-(2n+1)x}}\,,
\end{equation}
with $a_{0,0}=C$, $a_{1,0}=-\tfrac{1}{2}C$, $a_{2,0}=\tfrac{3}{8}C$, $a_{2,1}=\tfrac{1}{8}C^3$, etc., where $C$ (which we shall from now on swap for $A_\infty\equiv C\sqrt{2/\pi}$, cf.~\cite{Malinsky:2026eux}) is another a-priori unknown parameter that can not be determined just from the local behaviour of $y$ at $x\to \infty$. Barring numerics, the only known way of getting analytic grip on $A_\infty$ and $B_\infty$ is ``sewing'' the local expansions~(\ref{eq:Frobeniusat0intro}) and~(\ref{eq:InfinityTransseriesintro}) in a region of their common validity. However, given the finite convergence radius of the former and the asymptotic nature of the latter, this can only be achievend with a limited precision (presuming optimal truncation of~(\ref{eq:Frobeniusat0intro})), that, on the other hand, may be boosted significantly by stitching several Taylor-series intermediaries in between the two.        

\begin{figure}[h]
  \includegraphics[width=0.42\textwidth]{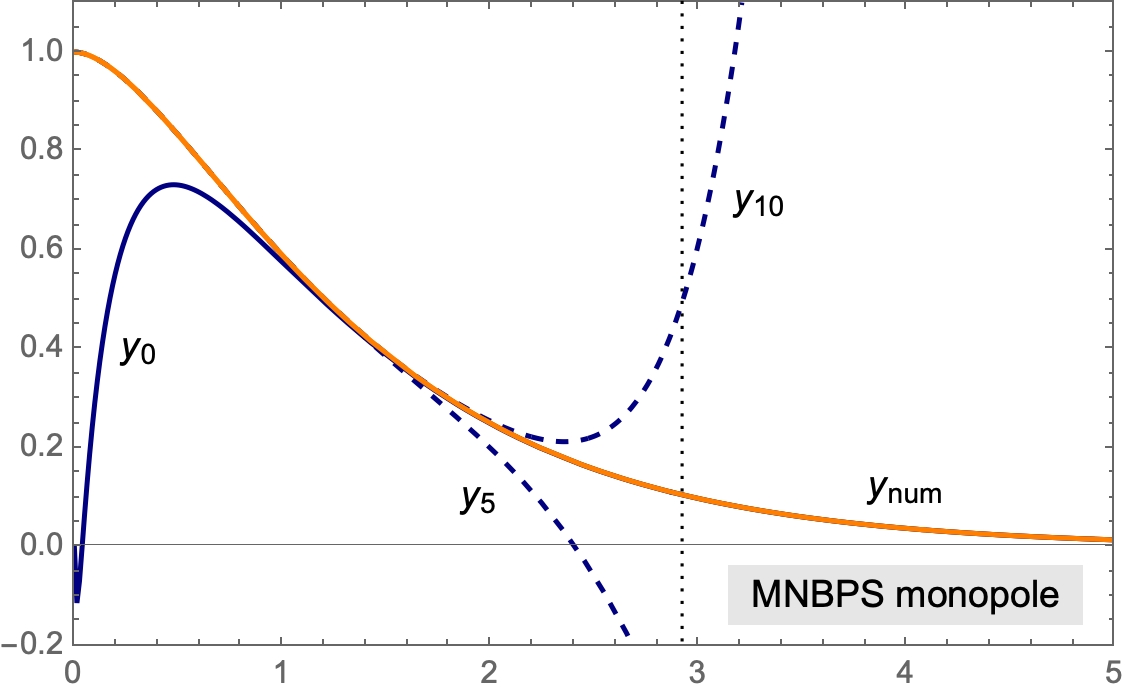}\;
\caption{\label{fig:combined}
The Borel-resummed $n=0$ tower of the asymptotic ($x\to\infty$) expansion~(\ref{eq:InfinityTransseriesintro}), corresponding to $y_0$ of    formula~(\ref{eq:universalasymptotics}) with $A_\infty=1.8$, is drawn in solid blue; the local ($x=0$) power-log expansion~(\ref{eq:Frobeniusat0intro}) with $B_\infty=-0.484$ truncated at two different values of $m_{\rm max}=5$ and $10$ ($y_5$ and $y_{10}$) is in dashed blue, and the ``exact'' numerical solution for the vector-profile ODE~(\ref{eq:MNBPSprofile}) of the MNBPS ($\beta\to\infty$) monopole is in orange. The vertical dashed line corresponds to the approximate radius of convergence of~(\ref{eq:Frobeniusat0intro}), $R\lesssim 2.92$, see also Appendix~\ref{app:MNBPSzeroprofile}. One can see a nice overlap of the numerical solution with the asymptotic background $y_0$ and also with the local truncated expansions $y_5$ and $y_{10}$ all the way to about $x\sim 1.4$.} 
\end{figure}

Recently, a new twist to this story was introduced in Ref.~\cite{Malinsky:2026eux} where the system of ODE's~(\ref{eq:gaugeprofile})-(\ref{eq:scalarprofile}) has been studied from the perspective of the resurgence theory (cf.~\cite{Ecalle:1981,Sternin:1996,Costin:2008,Aniceto:2011nu,Dunne:2013ada,marinoResurgenceNotes,Aniceto:2013fka,Dorigoni:2014hea,Sauzin:2016,Aniceto:2018bis,Dunne:2025mye}), exploiting the fact that, in the MNBPS case, the $n=0$ tower of the asymptotic expansion~(\ref{eq:InfinityTransseriesintro}) can be Borel-resummed into 
\be\label{eq:universalasymptotics}
y^\infty_0(x)=A_\infty \sqrt{x}K_{i\sqrt{3}/2}(x)\,.
\ee
This, in turn, had been contemplated in~\cite{Malinsky:2026eux} as a possible analytic background of a perturbative expansion of the full MNBPS solution; its performance as an analytic approximant of the ``exact'' numerical solution is illustrated in Fig.~\ref{fig:combined}.
Moreover, it was argued that this structure is fully universal in the sense that the same formula, albeit with a different numerical prefactor $A_\infty\to A_\beta$, applies to any NBPS resummed $n=0$ tower with arbitrary $\beta>0$.  
       
This observation has been recently used in the study~\cite{Dunne:2026hfx} where the authors developed an $x$-plane perturbative expansion around the universal asymptotic profile~(\ref{eq:universalasymptotics}), which greatly improved the quality of the analytic account of numerical solutions of Eqs.~(\ref{eq:gaugeprofile})-(\ref{eq:scalarprofile}), especially in the large-$x$ regime. For small $x$, the precision of the same expansion was not improving though, owing to the asymptotic nature of the profile~(\ref{eq:universalasymptotics}), which for $x\to 0$ tends to zero and, as argued in ver.2 of Ref.~\cite{Malinsky:2026eux}, is thus not an optimal background for expansions around $x=0$.

In this study, we exploit another hint provided in letter~\cite{Malinsky:2026eux}, namely, the possibility to rearrange the general NBPS perturbative expansion based on the asymptotic profile~(\ref{eq:universalasymptotics}) in such a way that a {\em new  pair of universal, nonperturbative, analytic and global background profiles obeying  boundary conditions}~(\ref{eq:boundaryconditionsy})-(\ref{eq:boundaryconditionsz}) {\em on both sides}, namely,
\begin{align}
y^G_0(x)&=\frac{1}{2}\left[e^{x}\left(1-\frac{1}{x}\right){\rm Ei}(-x)+e^{-x}\left(1+\frac{1}{x}\right){\rm Ei}(x)\right],
\nonumber\\
\label{eq:seedsintro}
z^G_0(x)&=1+\frac{1}{x^2\beta}\left[-1+e^{-\sqrt{2\beta}x}\left(1+\sqrt{2\beta}x\right)\right]\,,
\end{align}
with
\begin{align}\label{eq:Eidef}
{\rm Ei}(x)\equiv-\int_{-x}^{+\infty}\frac{e^{-z}}{z}{\rm d}z\,,
\end{align}
is identified as a basis of a global functional expansion that may even  converge uniformly to the true solutions of Eqs.~(\ref{eq:gaugeprofile})-(\ref{eq:scalarprofile}) on the entire~$\mathbb{R}_0^+$ and for all $\beta>0$. As we shall see, the key role here is again played by the modified Bessel $K_{\nu}$ and $I_{\nu}$ functions, but this time by those with $\nu=3/2$; as similar to the $\nu=i\sqrt{3}/2$ case of~(\ref{eq:universalasymptotics}) as it looks, we shall see that the qualitative difference between the two schemes will actually be enormous.           

Remarkably enough, the simplicity of the universal global profiles~(\ref{eq:seedsintro}) is even further pronounced in the Borel plane. As it turns out, the Borel counterpart of $y_0^G$ of Eq.~(\ref{eq:seedsintro}) is closely related to an elementary function $t\,{\rm arctanh}\,t=\tfrac{1}{2}\,t\,[\log(1+t)-\log(1-t)]$, whilst the $z_0^G$ therein even relates to purely linear profiles in the relevant local coordinates. This indicates a particular suitability of the Borel-plane perspective for the 't~Hooft-Polyakov monopole profile studies.

The paper is organized as follows: In Section~\ref{sec:threesubstitutions} we start with a detailed discussion of the two Borel-plane expansions recently developed in Ref.~\cite{Malinsky:2026eux}: the first one based on the Borel resummation of the universal $n=0$ tower of the asymptotic expansion of the vector profile for any $\beta>0$, that leads to formula~(\ref{eq:universalasymptotics}), and the second, rearranged one, that provides the universal analytic background~(\ref{eq:seedsintro}) obeying the correct boundary conditions both at zero and at infinity, as described above. 
These findings will, in turn, suggest a third, essentially trivial (albeit somewhat counterintuitive) substitution facilitating a simple {\em global, uniformly convergent, analytic approach} in the Laplace plane that, in Section~\ref{sec:MNBPS}, we shall first work out for the MNBPS ($\beta\to \infty$) setting, to be later (in Section~\ref{sec:NBPS}) generalized to an arbitrary NBPS ($\beta>0$) case. Yet another way of looking at the same problem from the Borel-plane perspective, arguably the most suitable for any beyond-leading-order analysis therein, is briefly commented upon in Section~\ref{sec:fourthsubstitution}. Then we conclude. Most of the technical details encountered along this route, including full  information on the expansions~(\ref{eq:Frobeniusat0intro}) and~(\ref{eq:InfinityTransseriesintro}), are deferred to a set of Appendixes.

\section{The three pictures/substitutions}\label{sec:threesubstitutions}

\subsection{Expanding around the asymptotic background\label{sec:asymptoticbackground}}

The shape of the universal {\em asymptotic background profile} for a general NBPS monopole identified in Ref.~\cite{Malinsky:2026eux},
\be\label{eq:asymptoticbackground}
y_0^\beta(x)=A_\beta \sqrt{x}K_{i\sqrt{3}/2}(x)\,,
\ee
closely resembling the asymptotics of the MNBPS monopole~(\ref{eq:universalasymptotics}), stems from the natural substitution in the scalar sector ODE~(\ref{eq:scalarprofile}),
\be\label{eq:subst1scalar}
\mbox{}{z(x)=1+g(x)}\,,
\ee
motivated by the fact that, for all $\beta>0$ (and even for $\beta=0$), $z(x)$ is in the vicinity of 1 on the vast majority of its domain, see Fig.~\ref{fig:numericsinintro}. With this at hand, the leading approximation to Eq.~(\ref{eq:gaugeprofile}) is nothing but the ODE~(\ref{eq:MNBPSprofile}), whose solutions' large-$x$ (asymptotic) expansion is discussed in detail in Appendix~\ref{app:MNBPSasymptoticprofile}. 

\subsubsection{Borel resummation of the $n=0$ asymptotic tower of~(\ref{eq:InfinityTransseriesintro})}
\begin{figure}[t]
  \includegraphics[width=0.42\textwidth]{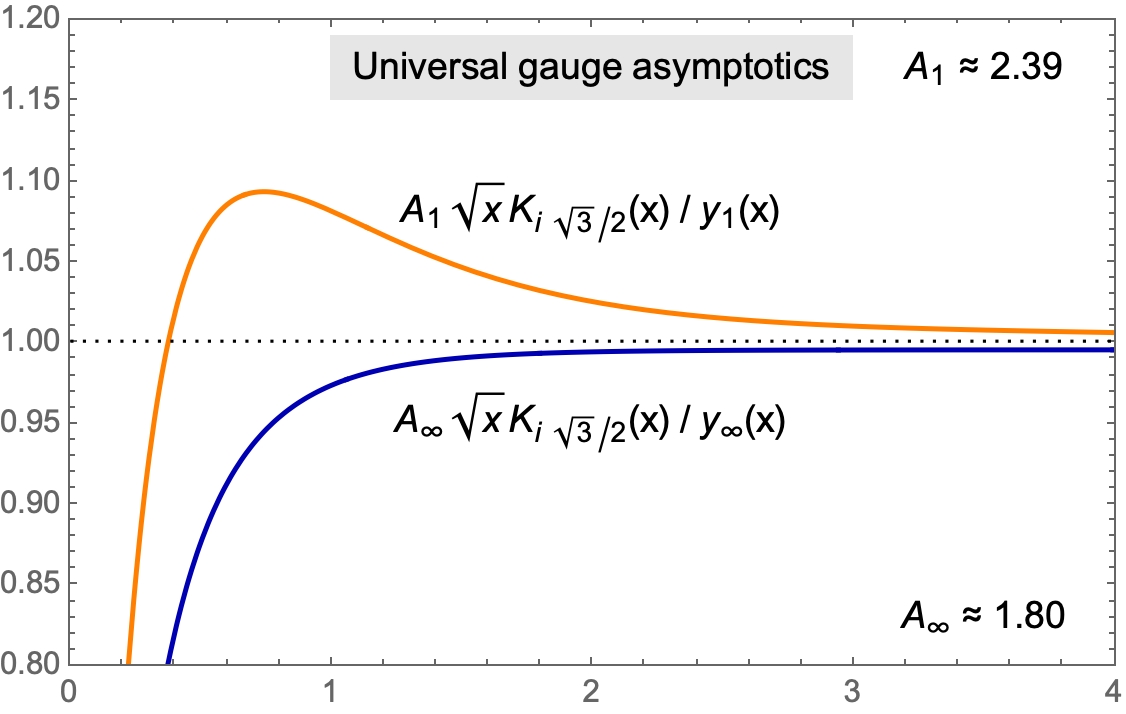}\;
\caption{\label{fig:yKasymptotics}
Ratios of the theoretical asymptotic gauge profiles~(\ref{eq:asymptoticbackground})  to the corresponding true numerical solutions (cf. Fig.~\ref{fig:numericsinintro}) for the MNBPS ($\beta\to\infty$) monopole (the lower curve, see also Fig.~\ref{fig:combined}), and for $\beta=1$ (the upper curve). Note that the asymptotics of the BPS solution ($\beta = 0$) is qualitatively different, falling at the leading order only as $2x e^{-x}$, see Eq.~(\ref{eq:BPS}).}  
\end{figure}

\begin{figure}[t]
  \includegraphics[width=0.42\textwidth]{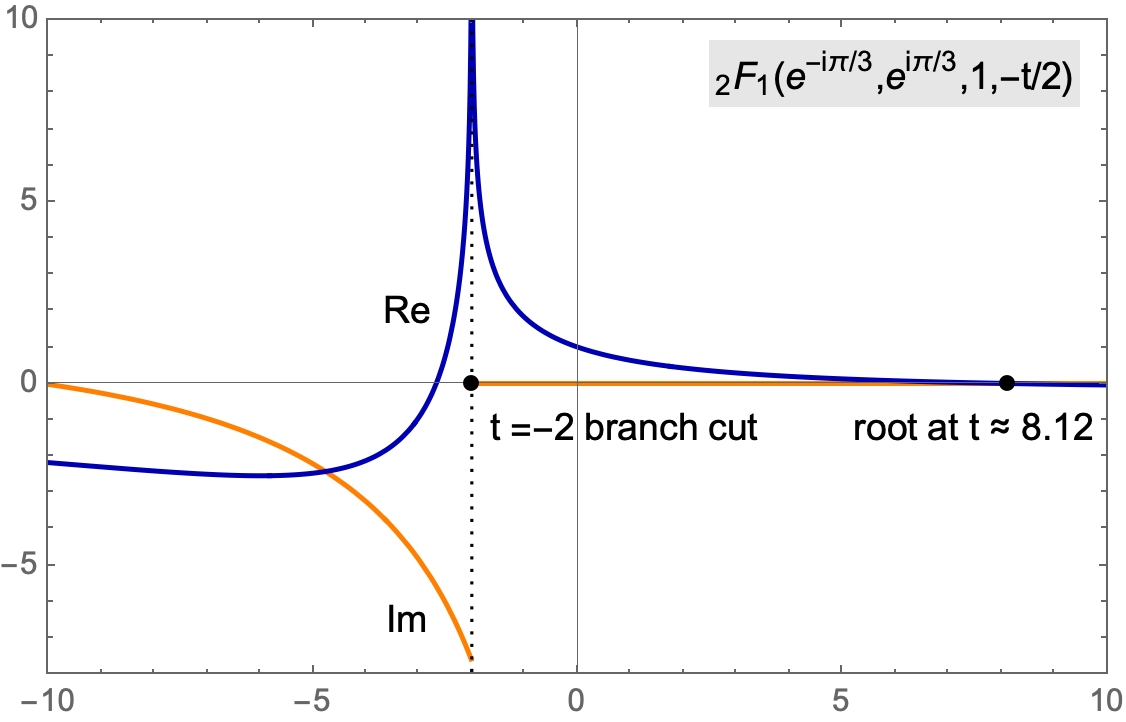}\;
\caption{\label{fig:Ffunction} The profile and important (principal branch) features of the ${}_2F_1(e^{-i\pi/3},e^{i\pi/3};1;-\tfrac{t}{2})$ Gaussian hypergeometric function governing the Borel-plane expansion around the asymptotic background $\hat{f}_0$ discussed in Sec.~\ref{sec:asymptoticbackground}.}
\end{figure}

From there and from Eq.~(\ref{eq:subst1scalar}) it follows that:
i) 
The ``linear'' $n=0$ tower of~(\ref{eq:InfinityTransseriesintro}) is completely universal, i.e. the same for all $\beta>0$, up to the unknown constant $C$ governing the $a_{m,0}$ numerical coefficients of  Eq.~(\ref{eq:amnnumericaltowers}), rewritten in Ref.~\cite{Malinsky:2026eux} as $A_\beta\sqrt{\frac{\pi}{2}}$; 
ii)
The same tower is most naturally Borel-resummed in coordinates corresponding to the gauge-sector substitution, cf.~(\ref{eq:y0intermsoff0}), 
\be\label{eq:subst1vector}
\mbox{}{y(x)=x e^{-x} f(x)}\,.
\ee
Hence, by virtue of~(\ref{eq:cm1intermsofPochhammer}), (\ref{eq:hypergeometricsindm}), (\ref{eq:dmcoefficients}) and (\ref{eq:f0resummation}), one obtains
\be\label{eq:y0intermsoff0xplane}
y_0(x)=x e^{-x}f_0(x)
\ee
where $f_0$ is a Laplace transform (in the regular germ)
\begin{equation}\label{eq:Laplacef0asymptotics}
f_0(x)=\int_0^\infty\!\! e^{-x t} \hat{f}_0(t)\equiv {\cal L}_t[\hat{f}_0(t)](x)
\end{equation}
of the Borel-plane structure 
\be\label{eq:hatf0tasymptotic}
\hat{f}_0(t)\equiv A_\beta\sqrt{\tfrac{\pi}{2}}\,{}_2F_1(e^{-i\pi/3},e^{i\pi/3};1;-\tfrac{t}{2})\,.
\ee
Taking into account (see for instance~\cite{Dunne:2025mye}) 
\begin{equation}\nn
\sqrt{x}K_\nu(x)=\sqrt{\tfrac{\pi}{2}}\,x e^{-x}\int_0^\infty\!\! e^{-x t}{}_2F_1(\tfrac{1}{2}-\nu,\tfrac{1}{2}+\nu;1;-\tfrac{t}{2})\,,
\end{equation}
one can, indeed, see that 
Eq.~(\ref{eq:asymptoticbackground}) is completely equivalent to Eq.~(\ref{eq:universalasymptotics}), just adapted for general $\beta>0$. 
Its universality is demonstrated in Fig.~\ref{fig:yKasymptotics}, and the salient features of the hypergeometric ``core'' of $\hat{f}_0$ are shown in Fig.~\ref{fig:Ffunction}.

One more point is perhaps worh making here: In a certain sense, the structure~(\ref{eq:y0intermsoff0xplane}) can also be viewed as a simple ``first-look'' dressing of a purely exponential background corresponding to the solution of Eq.~(\ref{eq:MNBPSprofile}) stripped from the entire $(y^3-y)/x^2$ nonlinearity, i.e., $y''=y$, which would yield just $e^{-x}$.  
\subsubsection{Laplace-plane perturbative expansion around $f_0$\label{sec:expansionaroundf0}}
In the coordinates of Eqs.~(\ref{eq:subst1scalar}) and~(\ref{eq:subst1vector}), the ODE system~(\ref{eq:gaugeprofile})-(\ref{eq:scalarprofile}) can be rewritten in the schematic form
\begin{align}
\label{eq:ODE'ssubstA}
L_A^f[f]&=e^{-2x}f^3+ g(g+2)f\\
L_A^g[g]&=2e^{-2x}(g+1)f^2+\beta g^2(g+3)\nn
\end{align} 
where the linear operators on the LHS's are 
\begin{align}
L_A^f[f](x)&\equiv f''(x)- 2 f'(x)+\frac{2f'(x)}{x}-\frac{2f(x)}{x}+\frac{f(x)}{x^2}\,,\nn\\
\label{eq:linearoperatorsA}
L_A^g[g](x)&\equiv g''(x)+\frac{2 g'(x)}{x}-2\beta g(x)  \,, 
\end{align}
and the RHS's correspond to non-linear forcing. In the MNBPS case, the same system reduces to $g(x)=0$ and
\begin{align}
L_A^f[f]&=e^{-2x}f^3\,.
\end{align}
It can be checked readily that the background profile $f_0$ of Eq.~(\ref{eq:Laplacef0asymptotics}) is a zero-mode of the operator $L_A^f$ and, hence, may serve as a background for a triangular expansion
\begin{align}
\label{eq:seriesfg}
f&=f_0+f_1+f_2+\ldots\\
g&=g_0+g_1+g_2+\ldots\nn
\end{align}
with $g_0=0$ and 
\begin{align}
L_A^f[f_0]&=0\,,\\
L_A^f[f_1]&=e^{-2x}f^3_0\,,\quad
L_A^g[g_1]=2e^{-2x}f^2_0\,,\nn
\\
L_A^f[f_2]&=3e^{-2x}f^2_0f_1+2f_0g_1\,,\quad
L_A^g[g_2]=2e^{-2x}g_1f^2_0\,,\nn
\end{align}
and so on, with the RHS's at each level corresponding to suitable subsectors of partitions\footnote{The rule of thumb here is that the sum of indices on the RHS is one less than the index on the LHS.} of the RHS's of EQs~(\ref{eq:ODE'ssubstA}) in terms of $f_i$'s and $g_j$'s. It is worth noting that the leading-order contribution to $g$ is forced by $f_0$ and, in this sense, the scalar sector is in fact {\em enslaved} by the gauge one, feeding back to $f$ only at the $f_2$ level.  
\subsubsection{Borel-plane perturbative expansion around $\hat{f}_0$\label{sec:MNBPSBorelPlane}}
As already shown in Ref.~\cite{Malinsky:2026eux}, the Borel-plane equivalent of Eqs.~(\ref{eq:ODE'ssubstA}) for $f$ and $g$ written in terms of Volterra-type equations for their $\hat{f}$ and $\hat{g}$ counterparts, reads 
\begin{align}
\label{eq:Volterraf}
&t(t+2)\hat{f}(t)+\int_0^t{\rm d}s K_A^f(t,s)\hat{f}(s)=\\
&\qquad\qquad\quad[\hat{f}*\hat{f}*\hat{f}](t-2)+2[\hat{f}*\hat{g}](t)+[\hat{f}*\hat{g}*\hat{g}](t)\,,\nn\\
\label{eq:Volterrag}
&(t^2-2\beta)\hat{g}(t)+\int_0^t{\rm d}s K_A^g(t,s)\hat{g}(s)=\\
&\qquad\; 2[\hat{f}*\hat{f}+\hat{f}*\hat{f}*\hat{g}](t-2)+\beta[3\,\hat{g}*\hat{g}+\hat{g}*\hat{g}*\hat{g}](t) \,,\nn
\end{align}  
where the two Volterra kernels therein given by
\begin{align}\label{eq:VolterrakernelsA}
K_A^f(t,s)\equiv t-3s-2\,,\quad
K_A^g(t,s)\equiv -2s\,,
\end{align}
{\em both depend on the integration variable} $s$. While this is not a problem for the general discussion of proliferation of the $t=-2$ ``seed singularity'' of $\hat{f}_0$ of Eq.~(\ref{eq:hatf0tasymptotic}) over the Borel plane, eventually (cf. Ref.~\cite{Malinsky:2026eux}) populating all even integers, it complicates practical calculations considerably, because the two equations~(\ref{eq:Volterraf})-(\ref{eq:Volterrag}) can not be simplified much, except for the MNBPS limit, where $\hat{g}(t)=0$ and Eq.~(\ref{eq:Volterraf}) assumes the form
\begin{align}
&t(t+2)\hat{f}(t)+\int_0^t{\rm d}s K_A^f(t,s)\hat{f}(s)=[\hat{f}*\hat{f}*\hat{f}](t-2)\nn\,.
\end{align}  

\subsubsection{Inadequacy of the asymptotic background $f_0$ for small $x$\label{sec:inadequacy}}
\begin{figure}[t]
  \includegraphics[width=0.42\textwidth]{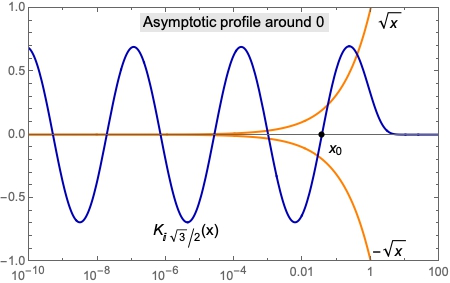}\;
\caption{\label{fig:asymptoticprofileatzero} The critical behaviour of the two components of the universal asymptotic profile~(\ref{eq:asymptoticbackground}) near $x=0$ (note the log scale of the horizontal axis). In blue, the rapid oscillation of its $K$-function part; in orange the $\pm\sqrt{x}$ envelope pushing the amplitude of $y_0$ towards zero for small $x$. As we argue in Sec.~\ref{sec:inadequacy}, this is the reason that $y_0^\beta$, whilst being a perfect expansion background at infinity, is not adequate in the small-$x$ regime. Note also the position of the rightmost root of $K$ at $x_0\sim 0.0373$, which will be needed in Sec.~\ref{sec:Ainftyparameter}.}
\end{figure}
As we already mentioned, the expansion developed around the asymptotic background $y_0^\beta$ of Eq.~(\ref{eq:asymptoticbackground}) (or around $\hat{f}_0$ in the Borel plane) is perfect for studying Eqs.~(\ref{eq:gaugeprofile})-(\ref{eq:scalarprofile}) in the asymptotic regime, i.e., for large $x$, see e.g.~\cite{Dunne:2026hfx}. However, its wild behaviour in the small-$x$ region, cf. Fig.~\ref{fig:asymptoticprofileatzero}, is such that the desired $y|_{x\to 0}\to 1$ boundary condition is hardly attainable at any perturbative order in this scheme. The reason is most easily seen in the MNBPS setting fully described by Eq.~(\ref{eq:MNBPSprofile}), whose singularity structure   admits only three {\em discrete} types of regular behaviour around $0$, namely,  $y\to\pm 1$ and $y\to 0$. These, however, are qualitatively different sectors that are not necessarily perturbatively connected, and, hence, the transition from the obvious $y|_{x\to 0}\to 0$ behaviour of each $y_0^\beta$ to the desired $y|_{x\to 0}\to 1$ of the physical solutions most likely requires a fundamental structural change, such as the one pointed out in Ref.~\cite{Malinsky:2026eux} and discussed in more detail the next subsection.    

\subsection{Universal global resummed background\label{sec:fRgRapproach}}
Remarkably enough, the problem elaborated on in Sec.~\ref{sec:asymptoticbackground} in terms of the asymptotic background functions $(f_0, g_0)$, with $g_0=0$ and $f_0$ being (in the Borel plane) the hypergeometric structure~(\ref{eq:hatf0tasymptotic}), may be reformulated  as a perturbative expansion around a {\em different non-perturbative background} $(f_0^R,\,g_0^R)$ which can be regarded as a rearrangement and partial resummation (${}^R$) of the two series~(\ref{eq:seriesfg}). This can be done in such a way that the corresponding ``seed profiles'' $y_0^R$ and $z_0^R$ (where the latter now likely becomes non-trivial) would obey the boundary conditions~(\ref{eq:boundaryconditionsy})-(\ref{eq:boundaryconditionsz}) both at $x=0$ as well as at $x\to \infty$. 
This, in turn, admits all higher-order corrections to tend to zero in both these limits  which, due to continuity, provides much more control over the ``overall size'' of the corresponding higher order contributions. As we shall see,  $f_0^R$ and $g_0^R$ will actually be nothing but the pair of {\em universal global non-perturbative background profiles} of Eq.~(\ref{eq:seedsintro}) that support a perturbative expansion satisfying the boundary conditions~(\ref{eq:boundaryconditionsy})-(\ref{eq:boundaryconditionsz}) at every perturbative levell, which shall {\em converge uniformly} to the corresponding exact solutions of Eqs.~(\ref{eq:gaugeprofile})-(\ref{eq:scalarprofile}). 
Technically, the key to this is a second substitution in the gauge sector 
\begin{align}
\label{eq:subst2vector}
f(x)&=f^R(x)+\frac{e^x}{x}\;,
\end{align}
along with a mere scalar sector coordinate relabelling
\begin{align}       
\label{eq:subst2scalar}
g(x)&=g^R(x)\;,
\end{align}
which, however, does not mean that the corresponding scalar background profile $g_0^R$ will be identical to $g_0=0$. 
As we shall see, this change of coordinates will bring some welcome features namely to the Borel-plane account of the emerging perturbative expansion; among other things, the ``rearranged/resummed'' variants of the Volterra kernels~(\ref{eq:VolterrakernelsA}) will become $s$-independent, simplifying the Borel picture considerably.      
 
\subsubsection{Laplace-plane expansion in the $f^R$, $g^R$ coordinates}
In terms of $f^R$ and $g^R$ defined by Eqs.~(\ref{eq:subst2vector}) and~(\ref{eq:subst2scalar}), the ODE system~(\ref{eq:gaugeprofile})-(\ref{eq:scalarprofile}) can be rewritten in the schematic form
\begin{align}
\label{eq:ODE'ssubstB1}
L_B^f[f^R]&=\frac{e^x}{x}+e^{-2x}(f^R)^3+3e^{-x}\frac{(f^R)^2}{x}
\\
&+\left(f^R+\frac{e^x}{x}\right)g^R(2+g^R)\,,\nn\\
\label{eq:ODE'ssubstB2}
L_B^g[g^R]&=2e^{-2x}(g^R+1)\left(f^R+\frac{e^x}{x}\right)^2\\
&+\beta (g^R)^2(g^R+3)\,,\nn
\end{align} 
where the linear operators on the LHS's read 
\begin{align}
L_B^f[f](x)&\equiv f''(x)- 2 f'(x)+\frac{2f'(x)}{x}-\frac{2f(x)}{x}-\frac{2f(x)}{x^2}\,,\nn\\
\label{eq:linearoperatorsB}
L_B^g[g](x)&\equiv g''(x)+\frac{2 g'(x)}{x}-\frac{2g(x)}{x^2}-2\beta g(x)  \,, 
\end{align}
and the RHS's again provide the forcing. 
The background profiles $f_0^R$ and $g_0^R$ around which one can, at least in principle, perturbatively expand the (\ref{eq:ODE'ssubstB1})-(\ref{eq:ODE'ssubstB2}) system, are in this case defined by
\begin{align}\label{eq:fRgRxplanedefs}
L_B^f[f_0^R]&=\frac{e^x}{x}\quad\text{and}\quad L_B^g[g_0^R]=\frac{2}{x^2}\,.
\end{align}
Note that, as anticipated, the scalar sector background $g_0^R$ can no longe be trivial here! Yet, as one can infer already from Eq.~(\ref{eq:ODE'ssubstB2}), it still remains zero in the MNBPS limit
where the system~(\ref{eq:ODE'ssubstB1})-(\ref{eq:ODE'ssubstB2}) reduces to 
\begin{align}
L_B^f[f^R]&=\frac{e^x}{x}+e^{-2x}(f^R)^3+3e^{-x}\frac{(f^R)^2}{x}\,.
\end{align}
This will be confirmed explicitly by the Borel plane calculation of $f_0^R$ and $g_0^R$ in the general $\beta>0$ case that we shall turn to now.
\subsubsection{Borel-plane calculation of $\hat{f}^R_0$, $\hat{g}^R_0$\label{sec:BorelcalculationofhatfRhatgR}}
\begin{figure}[th]
  \includegraphics[width=0.42\textwidth]{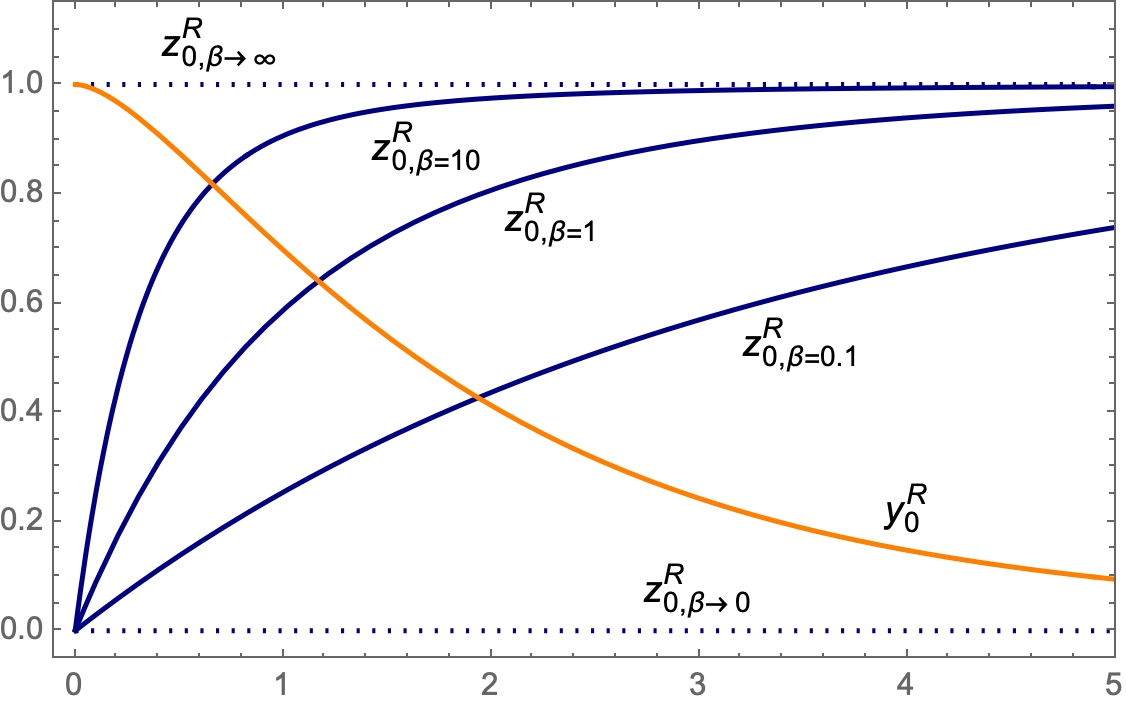}\;
\caption{\label{fig:fR0gR0profiles} The shapes of the resummed global background profiles $y_0^R$ and $z_0^R$ of Sec.~\ref{sec:fRgRapproach}. While $y_0^R$ (in orange) does not depend on $\beta$, $z_0^R$ does and it is shown for $\beta=0.1$, $1$ and $10$ (in blue), as well as in the MNBPS ($\beta\to\infty$) and the (formal) NBPS ($\beta\to 0$) limits (dotted blue). Note that for all $\beta>0$, these profiles can be used as global perturbative expansion backgrounds as explained in Secs.~\ref{sec:fRgRapproach}, \ref{sec:MNBPS} and~\ref{sec:NBPS}.}
\end{figure}
Due to the rather special shape of the linear operators~(\ref{eq:linearoperatorsB}) that differ from those of Eq.~(\ref{eq:linearoperatorsA}) in the $1/x^2$-proportional pieces, the Borel-plane counterparts of Eqs.~(\ref{eq:ODE'ssubstB1}),  (\ref{eq:ODE'ssubstB2}) assume particularly nice forms (with the generic $\hat{r}^f$, $\hat{r}^g$ symbols denoting the relevant RHS expressions)
\begin{align}
t(t+2)\hat{f}^R(t)+\int_0^t{\rm d}s K_B^f(t,s)\hat{f}^R(s)=\hat{r}^f(t)\,,\\
(t^2-2\beta)\hat{g}^R(t)+\int_0^t{\rm d}s K_B^g(t,s)\hat{g}^R(s)=\hat{r}^g(t)\,,
\end{align}  
because, by virtue of substitution~(\ref{eq:subst2vector}), both Volterra kernels therein,
\begin{align}
K_B^f(t,s)=-2(t+1)\,,\quad
K_B^g(t,s)=-2t\,,
\end{align}
become integration-variable ($s$) independent! This is extremely welcome because in such a case both these equations are equivalent to 1st order ODE's in
\begin{align}
F(t)\equiv\int_0^t{\rm d}s \hat{f}^R(s)\,,\quad G(t)\equiv\int_0^t{\rm d}s \hat{g}^R(s)\,,
\end{align}
reading
\begin{align}
t(t+2)F'(t)-2(t+1)F(t)=\hat{r}^f(t)\,,\\
(t^2-2\beta)G'(t)-2 t\, G(t)=\hat{r}^g(t)\,.
\end{align}  
These are easily solved by standard methods, providing
\begin{align}
\hat{f}^R(t)&=2C^f(t+1)\!+\!\frac{\hat{r}^f(t)}{t(t+2)}\!+\!2(t+1)\!\int_{t_0}^t\!{\rm d}s\frac{\hat{r}^f(t)}{s^2(s+2)^2}\,,\nn\\
\label{eq:generalhatfRgR}
\hat{g}^R(t)&=2C^g t+\frac{t\, \hat{r}^g(t)}{(t^2-2\beta)}+2t\int_{t_0}^t{\rm d}s\frac{s\, \hat{r}^g(s)}{(s^2-2\beta)^2}\,,
\end{align} 
see Appendixes~\ref{app:fundamentalODE} and~\ref{app:fundamentalODE'scalar}. For the specific RHS forms
\be
\hat{r}^f(t)=1_{-1}\,,\; \hat{r}^g(t)=2 t\,,
\ee
(notice the non-trivial germ of the map in $\hat{r}^f(t)$)
corresponding to the Borel-plane variant of the RHS's of the  defining equations~(\ref{eq:fRgRxplanedefs}), one reveals
\be\label{eq:hatfRsolution}
\hat{f}_0^R=(-1+u\, {\rm arctanh}\, u)_{-1}\,,
\ee
anchored at $-1$ (with local coordinate $u$). Similarly,
\be\label{eq:hatgRsolution}
\hat{g}_0^R=\left[\frac{t}{\beta}\right]_0\oplus \left[-\frac{1}{\beta}(v+\sqrt{2\beta})\right]_{\sqrt{2\beta}}\,,
\ee
with the first term written in the regular-germ coordinate $t$, whereas the second in the local coordinate $v$ of the germ at $\sqrt{2\beta}$. Note that the coefficient of the possible contribution from the $-\sqrt{2\beta}$ germ is zero; this is because the integration constants $C^f$ and $C^g$ (conspiring with the lower integration limits $t_0$ in formulae~(\ref{eq:generalhatfRgR})) had to been fixed in such a way that the appropriate\footnote{Note that these are, technically, evaluated as a median resummation, i.e., as an arithmetic average of a pair of lateral Laplace transforms avoiding, in a controlled manner, the real-axis singularities.}  Laplace transforms of expressions~(\ref{eq:hatfRsolution}) and~(\ref{eq:hatgRsolution}) 
\begin{align}
f^R_0(x)&=\frac{1}{2 x^2}\left[e^{2x}(x-1){\rm Ei}(-x)+(x+1){\rm Ei}(x)\right]-\frac{e^x}{x}\,,\nn\\
g^R_0(x)&=\frac{1}{\beta x^2}\left[-1+\exp(-\sqrt{2\beta}x)(1+\sqrt{2\beta}x)\right]\,, \label{eq:hatfRhatgRLaplaceTransforms}
\end{align} 
obeyed the boundary conditions 
\begin{align}
&x e^{-x}f_0^R(x)|_{x\to 0}\to 0\,,\quad  
x e^{-x}f_0^R(x)|_{x\to \infty} \to  -1\,, \nn\\
&g^R({x\to 0})\to -1\,,\; \qquad
g^R({x\to \infty}) \to  0\,, \nn
\end{align}
necessary to produce the desired behaviour of $y_0^R=1+x e^{-x}f_0^R$ and $z_0^R=1+g_0^R$. 

Notice that {\em no undetermined local constants} remain in $f^R$ or $g^R$ except for $\beta$ in the latter which, however, is an external parameter fixed by the physical setting, i.e., a constant of a different nature than, e.g., $A_\infty$ or $B_\infty$ in the local expansions~(\ref{eq:Frobeniusat0intro}) or~(\ref{eq:InfinityTransseriesintro}).
The shapes of thus obtained $y_0^R$ and $z_0^R$ profiles, which are nothing but the universal ``global backgrounds'' $y^G$ and $z^G$ of Eq.~(\ref{eq:seedsintro}) announced in Sec.~\ref{sec:introduction}, are illustrated in Fig.~\ref{fig:fR0gR0profiles} for a handful of sample values of $\beta$.      

Several comments are perhaps worth making here: 
i)
Around $x=0$, $y_0^R$ and $z_0^R$ behave like 
\begin{align}
\label{eq:y0Rexpansionatzero}
y^R_0(x)&=1+\tfrac{1}{3}x^2(\gamma_E-\tfrac{4}{3}+\log x)+{\cal O}(x^4)\,,\\
z^R_0(x)&=\tfrac{2}{3}\sqrt{2\beta}x-\tfrac{1}{2}\beta x^2+{\cal O}(x^3)\,,
\end{align}
where $\gamma_E$ is the Euler-Mascheroni constant. Note that the expression on the RHS of~(\ref{eq:y0Rexpansionatzero}) matches perfectly the local power-log expansion~(\ref{eq:Frobeniusat0intro}) (see also Appendix~\ref{app:MNBPSzeroprofile}).
ii)
Asymptotically, however, $y_0^R$ falls only as $2 x^{-2}$ and, hence, the exponential tails of the asymptotic expansions of ``true'' solutions (such as Eq.~(\ref{eq:InfinityTransseriesintro}) for the MNBPS profile) are fully attained only at the level of the full expansion tower. This, however, is not a real problem in ``traditional calculations'' based on the information about the true profiles (like the calculation of the monopole mass or determination of the asymptotic expansion parameters such as $A_\beta$ of Sec.~\ref{sec:threesubstitutions}) because the differences of the true exponential tails and those generated from $y_0^R$ at any given expansion order shall integrate to minuscule contributions that will tend to diminish at yet higher orders, see Sec.~\ref{sec:MNBPS}. 
iii)
The $z_0^R$ background as a function of $\beta$ behaves exactly as expected: for $\beta\to \infty$, the resummed scalar profile function assumes $g^R(x)=0$ which is what the MNBPS limit requires. 
iv)
In the $\beta\to 0$ limit $z_0^R$ formally vanishes {\em everywhere}; this, in fact, is just another indication of a qualitative difference between the NBPS ($\beta>0$) and the BPS ($\beta=0$) regimes.  

\subsection{Universal global expansion in the $x$-plane \label{sec:offsetsinboth}}
The intriguing results of Sec.~\ref{sec:fRgRapproach} indicate and interesting twist in what one considers to be natural substitutions in the scalar and gauge sectors of the system~(\ref{eq:gaugeprofile})-(\ref{eq:scalarprofile}) subject to boundary conditions~(\ref{eq:boundaryconditionsy})-(\ref{eq:boundaryconditionsz}). As for the former, setting $z=1+g^R$ looks very natural as it makes $g^R$ naturally small ``almost everywhere'', see also Sec.~\ref{sec:asymptoticbackground}. However, the key substitution~(\ref{eq:subst2vector}) that lead to the desired non-perturbative change between the $y|_{x\to 0}\to 0$ and $y|_{x\to 0}\to 1$ sectors, effectively corresponds to $y(x)=x e^{-x}f^R(x)+1$. At fist glance, the $+1$ offset therein looks counterintuitive because, unlike $z$, $y$ is close to 1 ``almost nowhere'' in $\mathbb{R}_0^+$.    
 
Nevertheless, let us take this finding, entertained so far mainly in the Borel plane, cf. Sec.~\ref{sec:fRgRapproach}, at the face value, and attempt to understand it better from the Laplace-plane perspective.
\subsubsection{The two offsets} 
Hence, in what follows, the central quantities of our interest will be the {\em offsets} of the gauge/vector and scalar profiles from 1  defined by       
\begin{align}\label{eq:vsparametrization}
\mbox{}{y(x)=1+v(x)}\;,\\
\mbox{}{z(x)=1+s(x)}\;.\nn
\end{align}
In the scalar sector, $s$ is just a (natural) relabelling of $g$ and/or $g^R$ used in Sects~\ref{sec:asymptoticbackground} and~\ref{sec:fRgRapproach} and also in Ref.~\cite{Malinsky:2026eux}, while $v(x)=x e^{-x}f^R(x)$ represents a new coordinate stripped from all the unnecessary load related to the Borel-resummation considerations of namely Sec.~\ref{sec:asymptoticbackground}.  

The original ODE system~(\ref{eq:gaugeprofile})-(\ref{eq:scalarprofile}) in these coordinates reads 
\begin{align}
\label{eq:nBPSvectorprofileinv}
&L_C^v[v]=1+\frac{1}{x^2}\left(3v^2+v^3\right)+(1+v)(2s+s^2)\,,\\
\label{eq:nBPSscalarprofileins}
&L_C^s[s]=\frac{2}{x^2}+\beta(3s^2+s^3)+\frac{2}{x^2}v(2+v)(1+s)\,,
\end{align}
where $L_C^v$ and $L_C^s$ denote the linear differential operators therein
\begin{align}\label{eq:Coperators}
&L_C^v[v]\equiv v''-\left(1+\frac{2}{x^2}\right)v\,, \\
&L_C^s[s]\equiv s''+\frac{2s'}{x}-\left(2\beta+\frac{2}{x^2}\right)s\nn\,.
\end{align}
Note that, as expected, $L_C^s$ is structurally identical to its Sec.~\ref{sec:fRgRapproach} counterpart $L_B^g$, while $L_C^v[v]$ is much simpler than $L_B^g$ therein.
Let us also note that in the MNBPS limit $s(x)=0$ and Eq.~(\ref{eq:nBPSvectorprofileinv}) assumes a very simple form
\begin{align}
\label{eq:MNBPSvectorprofileinv}
&L_C^v[v]=1+\frac{1}{x^2}\left(3v^2+v^3\right)\,.
\end{align}
\subsubsection{The triangular expansion\label{eq:generalvsexpansion}}  
A natural triangular perturbative scheme then emerges from the expansion
\begin{align}
\label{eq:expansionvs}
v=v_0+v_1+v_2+\ldots\,,\\
s=s_0+s_1+s_2+\ldots\,,\nn
\end{align}
which, substituted into Eqs.~(\ref{eq:nBPSvectorprofileinv})-(\ref{eq:nBPSscalarprofileins}) yields a tower of ODE's for the individual levels $v_n$ and $s_n$,
\begin{align}\label{eq:ODEforvn}
v_n''-\left(1+\frac{2}{x^2}\right)v_n&=r^v_n\,,\\
\label{eq:ODEforsn}
s_n''+\frac{2s_n'}{x}-\left(2\beta+\frac{2}{x^2}\right)s_n&=r^s_n\,,
\end{align}
with completely universal LHS's; their RHS's correspond to suitable partitions of those of Eqs.~(\ref{eq:nBPSvectorprofileinv})-(\ref{eq:nBPSscalarprofileins}), e.g.,  
\begin{align}\label{eq:rvs0}
r^v_0 &=1\,,\qquad
r^s_0=\frac{2}{x^2}\,,
\\
r^v_1&=\frac{1}{x^2}\left(3v_0^2+v_0^3\right)+(1+v_0)(2s_0+s_0^2)\,,\nn\\
r^s_1&=\beta(3s_0^2+s_0^3)+\frac{2}{x^2}v_0(2+v_0)(1+s_0)\,,\nn
\end{align}
and so on\footnote{Again, the $n$'th-level RHS's contain all combinations of $v_i$ and $s_j$ with the sum of their indices being $n-1$.}. In parametrization~(\ref{eq:vsparametrization}), the boundary conditions~(\ref{eq:boundaryconditionsy})-(\ref{eq:boundaryconditionsz}) translate  into
\begin{align}
\label{eq:boundaryconditionvbeta0}
&v_0({x\to 0})\to 0\,,\quad v_0({x\to \infty})\to -1\,,\\
&v_n({x\to 0})\to 0\,,\quad v_n({x\to \infty})\to 0\,,\; \text{for}\; n>0\,,\nn
\end{align}
and 
\begin{align}
\label{eq:boundaryconditionsbeta0}
&s_0({x\to 0})\to -1\,,\quad s_0({x\to \infty})\to 0\,,\\
&s_n({x\to 0})\to 0\,,\quad\;\; s_n({x\to \infty})\to 0\,,\; \text{for}\; n>0\,.\nn
\end{align}    
Note that, for each level $n$, the $r^{v,s}_n$ structures on the RHS's of Eqs.~(\ref{eq:ODEforvn})-(\ref{eq:ODEforsn}) are fixed and known as they are constructed solely from lower levels, i.e., from $v_i$'s and $s_j$'s with $i,j<n$.

The solutions of the two types of linear second order ODE's (\ref{eq:ODEforvn})-(\ref{eq:ODEforsn}) with non-trivial fixed RHS's are studied in detail in Appendixes~\ref{app:fundamentalODE}and~\ref{app:fundamentalODE'scalar}, both from the $x$-plane and also from the Borel-plane perspective. In what follows, we shall use that apparatus to construct the first few iterations of this scheme, starting with the simplified MNBPS ($\beta\to\infty$) setting, and then also in the general NBPS ($\beta>0$) case. In passing, we shall be able to get an exact non-perturbative formula for the $B_\infty$ coefficient governing the local expansion~(\ref{eq:Frobeniusat0intro}) of the former, and demonstrate that our method does indeed provide a uniformly convergent expansion of the exact solutions of Eqs.~(\ref{eq:gaugeprofile})-(\ref{eq:scalarprofile}). 
      
\section{The MNBPS monopole}\label{sec:MNBPS}
In the MNBPS limit, $s_n(x)=0$ to all orders, and the vector-profile ODE~(\ref{eq:ODEforvn}) is equipped with a stream of $r_n^v$ RHS's generated solely by the RHS of the defining equation~(\ref{eq:MNBPSvectorprofileinv}); at the first few levels one obtains (dropping throughout this Section the unnecessary superscript $v$)
\begin{align}
\label{eq:r0}
r_0(x)&=1\,,\\
r_1(x)&=\frac{1}{x^2}\left(3v_0^2+v_0^3\right)\,,\nn\\
r_2(x)&=\frac{1}{x^2}\left(3v_0^2v_1+6v_0v_1\right)\,,\nn\\
r_3(x)&=\frac{1}{x^2}\left(3v_0^2v_2+6v_0v_2+3v_0v_1^2+3v_1^2\right)\,,\nn\\
\label{eq:r4}
r_4(x)&=\frac{1}{x^2}\left(3v_0^2v_3+6v_0v_3+6v_0v_1v_2+v_1^3+6v_1v_2\right)\,,\nn
\end{align}
and so on. 
%

\subsection{Leading-order ($n=0$) correction, MNBPS case \label{sec:zerothapproximation}} 
With this at hand, one can readily write the analytic form of $v_0$ defined by the ODE~(\ref{eq:ODEforvn}), with $r_0(x)=1$ on its RHS, and the boundary condition~(\ref{eq:boundaryconditionvbeta0}). Following standard methods recapitulated in Appendix~\ref{app:fundamentalODE}, the general solution of Eq.~(\ref{eq:ODEforvn}) reads\footnote{Due to limited space, we shall sometimes suppress the integration variables in the particular solution integrals.}
\be\nn
v_0(x)=C_{1,0} h_1(x)+C_{2,0} h_2(x)+h_1\int_{x_0}^x h_2-h_2\int_{x_0}^x h_1\,,
\ee 
where $h_1$ and $h_2$, the fundamental modes of the $L_C^v[v]$ operator of Eq.~(\ref{eq:Coperators}), are nothing but (conveniently normalized) modified Bessel $I_{3/2}$ and $K_{3/2}$ functions\footnote{Interestingly, almost the same basis (differing from ours just by overall $x^{-1}$ factors) has already been used in a similar context in~\cite{Bais:1976}, albeit with a different scope.} multiplied by $\sqrt{x}$, see their defining formulae~(\ref{eq:h1vector})-(\ref{eq:h2vector}). The   
boundary condition~(\ref{eq:boundaryconditionvbeta0}) then requires
\begin{align}
C_{1,0}&= -\int_{x_0}^\infty h_2(z)r_0(z){\rm d}z\,,\\
C_{2,0}&= \int_{x_0}^0 h_1(z)r_0(z){\rm d}z\,,
\end{align}
which enforces
\be\label{eq:v0formgeneral}
v_0(x)=-h_1\!\int_{x}^\infty \!\! h_2-h_2\!\int_{0}^x h_1=\int_0^\infty \!\!G(x,z){\rm d}z\,,
\ee 
where 
\begin{align}
G(x,z)\equiv -[h_1(x)h_2(z)\theta(z\!-\!x)\!+\!h_2(x)h_1(z)\theta(x\!-\!z)]
\label{eq:Greensfunction1}
\end{align} 
is the Green's function of the linear problem. 
Remarkably, the integration~(\ref{eq:v0formgeneral}) can be done analytically, 
\be\label{eq:v0analytic}
v_0(x)=-1+{\rm Ei}(-x)h_1(x)+{\rm Shi}(x)h_2(x)\,,
\ee 
with the Ei function defined by formula~(\ref{eq:Eidef}) and
\begin{align}
{\rm Shi}(x)&\equiv\int_{0}^{x}\frac{\sinh z}{z}{\rm d}z\,.
\end{align}
Note that, up to the first term therein, Eq.~(\ref{eq:v0analytic}) is just yet another way of writing $y_0^G$ of Eq.~(\ref{eq:seedsintro}), which is to be expected on the consistency grounds, see also Eq.~(\ref{eq:hatfRhatgRLaplaceTransforms}) of Sec.~\ref{sec:BorelcalculationofhatfRhatgR}.
Hence, all findings about the nice match of $y_0=1+v_0$ to the local transseries~(\ref{eq:Frobeniusat0intro}) as well as the form of the leading order contribution to $B_\infty$ obtained in Sec.~\ref{sec:BorelcalculationofhatfRhatgR} follow here.

\subsection{Higher-order ($n>0$) corrections, MNBPS case\label{sec:MNBPShigherorders}} 
\begin{figure}[t]
  \includegraphics[width=0.42\textwidth]{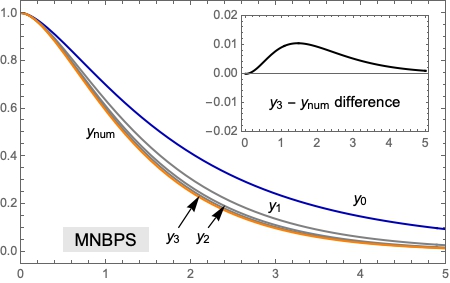}\;
\caption{\label{fig:y0y1y2}
The universal global non-perturbative gauge profile $y_0=1+v_0$, (also called $y^G$ in Sec.~\ref{sec:introduction}, in blue), and the first few perturbative approximations $y_1$, $y_2$ and $y_3$ (in gray) to the ``exact'' numerical MNBPS solution $y_{\rm num}$ (in orange). Note that the pace of the uniform global convergence of the $y_n$ sequence is such that $y_3$ is already barely distiguishable from $y_{\rm num}$; see also the inlay picture, Fig.~\ref{fig:RHSr1r2} and Sec.~\ref{sec:Binfinityexact}.} 
\end{figure}

\begin{figure}[t]
  \includegraphics[width=0.42\textwidth]{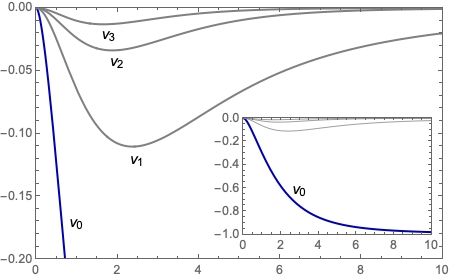}\;
\caption{\label{fig:v0v1}
The first few contributions to the full MNBPS profile function $y_\infty=1+\sum_{n\geq 0} v_n$ of Eq.~(\ref{eq:yndefinition}) ($v_0$ in blue and $v_1$, $v_2$, $v_3$ in gray) displayed over the (vertical) range $(0,-0.2)$, with the small inlay picture illustrating the global situation on the entire $(0,-1)$ range. One can notice the qualitative difference of the behaviour of $v_0$ defining the global analytic non-perturbative background, and that of the perturbative corrections $v_{n>0}$ fast diminishing in both small-$x$ and large-$x$ regimes. This behaviour is important for the geometric suppression of $r_n$'s of Eq.~(\ref{eq:r0}) for large $n$ and, hence, for the fast uniform convergence of the expansion.} 
\end{figure}

\begin{figure}[t]
  \includegraphics[width=0.42\textwidth]{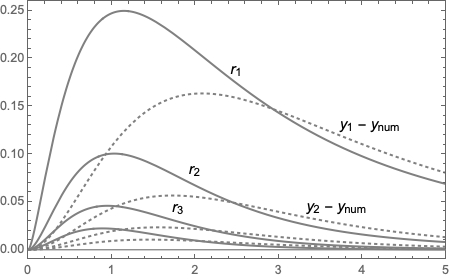}
\caption{\label{fig:RHSr1r2}
In solid gray, the RHS's $r_1$, $r_2$, $r_3$ and $r_4$ (cf. Eq.~(\ref{eq:r0})) of  ODE's~(\ref{eq:ODEforvn}) in case of the MNBPS monopole; in dotted gray, deviations of the $y_0$, .., $y_3$ functions of Eq.~(\ref{eq:yndefinition}) from the ``exact'' numerical profile, see also Figs.~\ref{fig:numericsinintro} and~\ref{fig:y0y1y2}. It is encouraging to see the $r_n$'s diminish uniformly and rapidly, remaining most prominent in the regions where they are needed most, i.e., where $y_n$'s most differ from the exact solution.} 
\end{figure}

Remarkably, the very same pattern is observed also at higher orders; the  $v_n$'s obeying the relevant boundary conditions~(\ref{eq:boundaryconditionvbeta0}) are {\em at all orders} given just by 
\begin{align}\label{eq:vnformgeneral}
v_n(x)&=-\left(h_1\int_{x}^\infty h_2 r_n +h_2\int_{0}^x h_1 r_n\right)\nn\\
&=\int_0^\infty G(x,z)r_n(z){\rm d}z\,.
\end{align}
For illustration, the shape of the first three approximations $y_0$, $y_1$, $y_2$ defined by   
\be\label{eq:yndefinition}
y_n=1+\sum_{m=0}^n v_m\, 
\ee
(with $y_0$ identical to $y^G_0$ of Eq.~(\ref{eq:seedsintro})), compared to the full numerical solution $y_{\rm num}$, is depicted in Fig.~\ref{fig:y0y1y2}, while the individual contributions 
$v_0$, $v_1$ and $v_2$ are shown in Fig.~\ref{fig:v0v1}. 
It is also interesting to see how nicely the supports of the forcing terms $r_n$ match those of the deviations $y_n-y_{\rm num}$, see Fig.~\ref{fig:RHSr1r2}.
From the same plot, one can also see the fast uniform convergence of $|r_n|$ to zero, as anticipated in Sec.~\ref{sec:uniformconvergence}. 
\subsection{Dyson equation interpretation} 
\noindent From~(\ref{eq:vnformgeneral}) one can see that the problem of finding the profile of the MNBPS monopole can be reformulated in terms of a Dyson expansion  
\begin{align}
v(x)&=\int_0^\infty G(x,z)\left[1+\frac{3v(z)^2+v(z)^3}{z^2}\right]{\rm d}z\,,
\end{align}
which is a prototype structure one can deal with by the standard methods of QFT perturbative expansion (including Feynman diagrams; see also~\cite{Bais:1976} and~\cite{Gardner:1983}). This is facilitated by the fact that the Green's function~(\ref{eq:Greensfunction1}) therein is known exactly,
cf. Eqs~(\ref{eq:fundB1})-(\ref{eq:fundB2}), 
\begin{align}
\label{eq:Greensfunction2}
G(x,z)= \sqrt{xz}&\left[-I_{3/2}(x)K_{3/2}(z)\theta(z-x)\right.\\
&\;\left.-K_{3/2}(x)I_{3/2}(z)\theta(x-z)\right]\,,\nn
\end{align} 
where $I_{3/2}$ and $K_{3/2}$ are the modified Bessel functions. 
A further exploitation of this fact is, however, beyond the scope of the current study and may be elaborated on elsewhere.
\subsection{Uniform convergence of the expansion\label{sec:uniformconvergence}} 
With this information at hand, one can already prove the uniform convergence of the expansion~(\ref{eq:expansionvs}). Two ingredients come to play here, namely, i) the {\em only polynomial growth of the number of terms at each level of the triangular expansion}~(\ref{eq:expansionvs}),  (roughly proportional to $n(n+1)/2$), cf.~(\ref{eq:r0}), with none of the coefficients therein ever exceeding $6$; and ii) the observation of the {\em geometrical boundedness of the global extrema of all $v_n$'s} following also from there (and clearly visible, e.g., in Fig.~\ref{fig:RHSr1r2}), i.e.,
\be
\max_{x\in \mathbb{R}_0^+}|v_n(x)|\lesssim c\,q^n \qquad \forall n\in \mathbb{N}\,,
\ee 
for a fixed real $c$ and the quotient $q$ of roughly $0.3$. Hence, the $\max_{x\in \mathbb{R}_0^+}|r_n|$ progression will behave asymptotically like $n^2 q^n$ which means that $|r_n|\rightrightarrows 0$ and, thus, $|v_n|\rightrightarrows 0$.
\subsection{The $B_\infty$ parameter to all orders\label{sec:Binfinityexact}} 
The $B_\infty$ parameter of the local expansion~(\ref{eq:Frobeniusat0intro}) is defined as the coefficient of the $x^2$ term therein that, owing to the analytic features of the $h_1$ and $h_2$ functions, can be at each perturbative ($n>0$) level exposed as
\be\label{eq:v1expansion}
v_n(x)= -\frac{1}{3}x^2\int_0^\infty h_2(z)r_n(z){\rm d}z+{\cal O}(x^3)\,.
\ee
Hence, $B_\infty$ can be written {\em in a closed form} 
\be\label{eq:Bfull}
B_\infty=B_{\infty,0}-\frac{1}{3}\int_0^\infty {\rm d}z\, h_2(z)\frac{3v^2+v^3}{z^2}\,,
\ee  
where the partition relation
\be
\sum_{n=1}^\infty r_n=\frac{1}{x^2}\left(3v^2+v^3\right)
\ee 
in the core of the perturbative approach to~(\ref{eq:MNBPSvectorprofileinv}) has been utilized, and the leading contribution,
\be\label{eq:Binfty0}
B_{\infty,0}=\tfrac{1}{3}(\gamma_E-\tfrac{4}{3})\approx -0.252\,,
\ee
has been extracted from the local expansion~(\ref{eq:y0Rexpansionatzero}).
Interestingly, this makes it possible to verify the validity of our findings by evaluating $B_\infty$ from~(\ref{eq:Bfull}) with the integral part calculated from the known numerical profile of the MNBPS monopole, cf. Fig.~\ref{fig:numericsinintro}. One thus obtains\footnote{Interestingly, using the very good approximate form of $v$, namely, $v_{\rm ap.}=2e^{-x}-e^{-2x}-1$, instead of the true numerical solution one obtains $B\sim -0.449$, quite off the target (although, optically, the two curves look rather similar, at least for small $x$).}
\be
B_\infty\approx -0.484\,,
\ee 
which is {\em spot on} at the fitted numerical value~(\ref{eq:Binftyfitted}) of Appendix~\ref{app:MNBPSzeroprofile}, see also Fig~\ref{fig:y0frobeniusexpansion}. Hence, the scheme developed in this work is fully consistent.

Finally, let us also note that the value of the RHS of Eq.~(\ref{eq:Bfull}) is very sensitive to the quality of a specific fit/approximation and, as such, it may be used as a simple measure of the same. For instance, one obtains, respectivelly, 
$B_{\infty,1}\approx -0.391$, 
$B_{\infty,2}\approx -0.444$, 
$B_{\infty,3}\approx -0.466$,
$B_{\infty,4}\approx -0.475$,
for $y_0$, $y_1$, $y_2$ and  $y_3$ (recall that this is enough to calculate $r_i$'s on the RHS of Eq.~(\ref{eq:v1expansion}) up to $r_4$) depicted in Fig.~\ref{fig:y0y1y2}. 
\subsection{The $A_\infty$ parameter\label{sec:Ainftyparameter}} 
\begin{figure}[t]
  \includegraphics[width=0.42\textwidth]{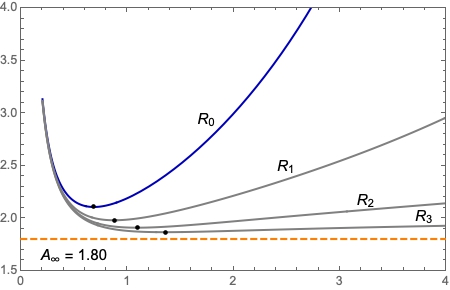}\;
\caption{\label{fig:Acoef}
The ratios $R_n(x)=y_n(x)/\sqrt{x}K_{i\sqrt{3}/2}(x)$ of Eq.~(\ref{eq:Ainftyformula}) for $n=0$ (in blue) and $n=1,2,3$ (in gray). One expects that for rising $n$ the sequence of the minima of the solid curves converges to $A_\infty\approx 1.8$ (dashed orange line), see also Eq.~(\ref{eq:universalasymptotics}). For the four indicated points one gets $R_0(0.68)\approx 2.10$, $R_1(0.87)\approx 1.98$, $R_2(1.09)\approx 1.91$ and $R_3(1.35)\approx 1.86$, which is encouraging.} 
\end{figure}

Interestingly, with all this at hand, one can also get a grip on the $A_\infty$ parameter of the asymptotic expansion~(\ref{eq:InfinityTransseriesintro}), albeit not as directly as it was the case for $B_\infty$ in Sec.~\ref{sec:Binfinityexact}. As was already mentioned in Sec.~\ref{sec:BorelcalculationofhatfRhatgR}, the tower of resurgent exponential tails of the full asymptotic solution, cf. Eq.~(\ref{eq:InfinityTransseriesintro}), is never fully reconstructed at finite $n$. Hence, it {\em does not} make sense to calculate $A_\infty$ as a limit of a sequence of terms (cf. Eq.~(\ref{eq:universalasymptotics}))
\be
\lim_{x\to \infty}y_n(x)/\sqrt{x}K_{i\sqrt{3}/2}(x) \,, 
\ee
because they all simply diverge at $\infty$, see Fig.~\ref{fig:Acoef}. Nevertheless, there is still a way to exploit the $y_n(x)/\sqrt{x}K_{i\sqrt{3}/2}(x)$ ratios to expose $A_\infty$ in a controlled manner; the trick is to look at the limit of the {\em sequence of minima} of these functions instead, 
\be\label{eq:Ainftyformula}
A_\infty=\lim_{n\to \infty}\min_{x>x_0}\left[y_n(x)/\sqrt{x}K_{i\sqrt{3}/2}(x)\right]\,,
\ee 
where $x_0\sim 0.0373$ is the rightmost root of $K_{i\sqrt{3}/2}$, cf. Fig.~\ref{fig:asymptoticprofileatzero}. Note that this is based on the observation that in the MNBPS case the asymptotic profile~(\ref{eq:universalasymptotics}) approaches the exact numerical solution monotonically, cf. Fig.~\ref{fig:yKasymptotics}. At $n=0,1,2$ and $3$, one thus obtains a sequence $A_{\infty,0}\approx 2.10$, $A_{\infty,1}\approx 1.98$, $A_{\infty,2}\approx 1.91$ and $A_{\infty,3}\approx 1.86$ which, eventually, should converge to the fitted numerical value $A_\infty\approx 1.80$. 

\section{The NBPS monopole}\label{sec:NBPS}
In the general $\beta>0$ case, one has to employ the full machinery of Sec.~\ref{eq:generalvsexpansion} in both vector and scalar sectors.  The  technical prerequisites of this undertaking are prepared in Appendixes~\ref{app:fundamentalODE} and~\ref{app:fundamentalODE'scalar}, along with the first two sets of RHS's~(\ref{eq:rvs0}) of the fundamental ODE's~(\ref{eq:ODEforvn})-(\ref{eq:ODEforsn}). 
\subsection{Leading-order ($n=0$) contribution, NBPS case \label{sec:zerothapproximationNBPS}}
Even for finite $\beta$, the shape of $v_0$ remains the same as for the MNBPS monopole, cf.~(\ref{eq:v0analytic}), which is just a demonstration of the anticipated universality of the global vector background profile $y^G_0$ of Eq.~(\ref{eq:seedsintro}). 

Concerning the scalar sector, the general form of $s_0$ is given by
\be\label{eq:v0form}
s_0(x)=C_{1,0} g_1(x)+C_{2,0}g_2(x)-g_1\int_{x_0}^x \frac{g_2}{W}r_0^s+g_2\int_{x_0}^x \frac{g_1}{W}r_0^s\,,
\ee 
where $g_1$ and $g_2$ are related to $h_1$ and $h_2$ of Eqs.~(\ref{eq:h1vector})-(\ref{eq:h2vector}) via
\begin{align}
g_1(x)=\frac{1}{x}h_1(\sqrt{2\beta}x)\,,\quad
g_2(x)=\frac{1}{x}h_2(\sqrt{2\beta}x)\,,
\end{align}
and the Wronskian $W$ which, due to the presence of a first derivative term in Eq.~(\ref{eq:ODEforsn}), is nontrivial here, reads
\be
W(g_1,g_2)=-\frac{\sqrt{2\beta}}{x^2}\,.
\ee
Imposing the boundary conditions~(\ref{eq:boundaryconditionsbeta0}) one arrives at
\be\label{eq:s0formgeneral}
s_0(x)=g_1\int_{x}^\infty \frac{g_2}{W} r_0^s+g_2\int_{0}^x \frac{g_1}{W} r_0^s\,,
\ee 
which, for $r_0^s=2/x^2$, is again a remarkably simple expression, structurally almost identical\footnote{Note that the apparently extra minus sign in formula~(\ref{eq:v0formgeneral}) is due to the Wronskian which, unlike here, had been explicitly evaluated there.} to that for $v_0$ (cf. Eq.~(\ref{eq:v0formgeneral})). Importantly, it can again be  written in a closed form
\begin{align}
\label{eq:seeds}
s_0(x)&=\frac{1}{x^2\beta}\left[-1+e^{-\sqrt{2\beta}x}\left(1+\sqrt{2\beta}x\right)\right]\,.
\end{align}
Hence, one confirms the Borel-plane result~(\ref{eq:hatfRhatgRLaplaceTransforms}) also from the $x$-plane perspective; needless to say, both these confirm the anticipated form of the universal global scalar background profile $z^G_0$ of Eq.~(\ref{eq:seedsintro}). 

\subsection{Higher-order ($n>0$) corrections, NBPS case}
Obviously, the algebraic account of higher order corrections in the general $\beta>0$ NBPS case is more tedious than that of the MNBPS limit of Sec.~\ref{sec:MNBPShigherorders}. 
Nevertheless, one still has
\be\label{eq:snformgeneral}
s_n(x)=g_1(x)\!\!\int_{x}^\infty \frac{g_2(z)}{W(z)} r_n^s(z){\rm d}z+g_2(x)\!\!\int_{0}^x \frac{g_1(z)}{W(z)} r_n^s(z){\rm d}z,
\ee 
and the main qualitative difference is the $\beta$-dependence of all the results that sneaks into the gauge sector already at the $n=1$ level via the $s_0$-proportional contribution present in $r_1^v$ of Eq.~(\ref{eq:rvs0}). 
Hence, we shall focus here mainly on the numerical illustration of our findings.    

\begin{figure}[t]
  \includegraphics[width=0.42\textwidth]{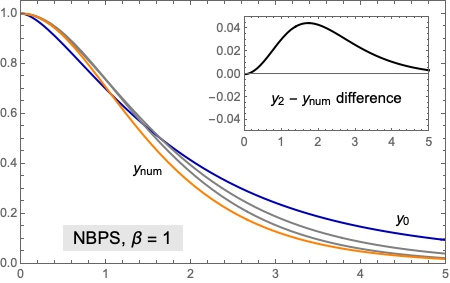}\;
  \includegraphics[width=0.42\textwidth]{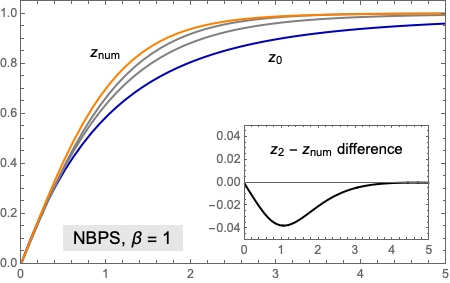}\;
\caption{\label{fig:y0y1z0z1beta1}
The gauge and scalar profiles of the $\beta=1$ monopole as expanded in Sec.~\ref{sec:NBPS}. The universal global backgrounds $y_0$, $z_0$ (also called $y^G_0$ and $z^G_0$ in Sec.~\ref{sec:introduction}, cf. Eqs.~(\ref{eq:seedsintro})) in blue, the first two perturbative orders  $y_1=y_0+v_1$, $y_2=y_0+v_1+v_2$ and $z_1=z_0+s_1$, $z_2=z_0+s_1+s_2$, respectively, in gray. The orange curves correspond to the numerical solutions $y_{\beta=1}$ and $z_{\beta=1}$ of~Fig.~\ref{fig:numericsinintro}. One can again appreciate the uniform convergence of the expansion, albeit slower than that in the MNBPS case of Sec.~\ref{sec:MNBPS}, cf. Fig.~\ref{fig:y0y1y2}.} 
\end{figure}

In Fig.~\ref{fig:y0y1z0z1beta1}, one can again appreciate a nice convergence of $y_n$ of Eq.~\ref{eq:yndefinition} and $z_n$ of 
\be
z_n=1+\sum_{m=0}^n s_m\, 
\ee
to the ``exact'' numerical profiles for $n=0$, $1$ and $2$ in case of the $\beta=1$ NBPS monopole. Note that the non-perturbative gauge background is exactly the same like that of the MNBPS case of Sec.~\ref{sec:MNBPS}, and also that in the current formalism all the differences between the MNBPS and NBPS solutions in the gauge sector come solely from the $\beta$-dependent perturbative corrections.   
\begin{figure}[t]
  \hskip -2mm\includegraphics[width=0.43\textwidth]{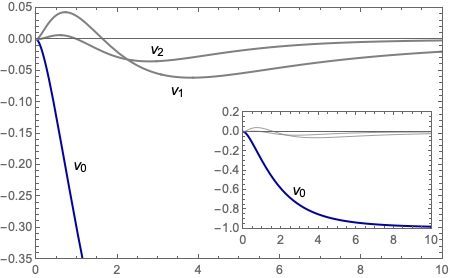}\;
  \includegraphics[width=0.42\textwidth]{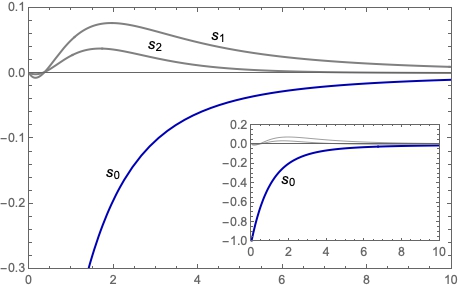}\;
\caption{\label{fig_v0v1v2beta1}
Demonstration of the gradual diminishing of perturbative corrections $v_1$ and $v_2$ in the gauge sector of the NBPS monopole with $\beta=1$ in the upper panel, and $s_1$ and  $s_2$ of the scalar sector in the lower panel, along with the nontrivial parts of the corresponding global analytic backgrounds ($v_0$ and $s_0$) in blue. Global comparison of the non-perturbative and perturbative sectors is depicted in the inlay panels.} 
\end{figure}

Finally, let us note that also in this case the problem may be formulated in terms of a Dyson system
\begin{align}
&v(x)\!=\!\int_0^\infty \!\!\!G(x,z)\!\!\left[1\!+\!\frac{1}{z^2}\left(3v^2\!+\!v^3\right)\!+\!(1\!+\!v)(2s\!+\!s^2)\right]{\rm d}z,\nn\\
&s(x)\!=\!\int_0^\infty \!\!\!\tilde{G}(x,z)\!\!\left[\frac{2}{z^2}\!+\!\beta(3s^2\!+\!s^3)+\frac{2}{z^2}v(2\!+\!v)(1\!+\!s)\right]{\rm d}z\nn,
\end{align}
with the gauge-sector Green's function $G$ of Eq.~(\ref{eq:Greensfunction1}) complemented with its scalar-sector counterpart
\begin{align}
\tilde{G}(x,z)&=-\frac{z}{2\beta x}[h_1(\sqrt{2\beta}x)h_2(\sqrt{2\beta}z)\theta(z-x)\!
\\
\nn
&\qquad\quad\; +\!h_2(\sqrt{2\beta}x)h_1(\sqrt{2\beta}z)\theta(x-z)]\,.
\end{align}

\section{The fourth substitution}\label{sec:fourthsubstitution}
\begin{figure}[t]
  \includegraphics[width=0.42\textwidth]{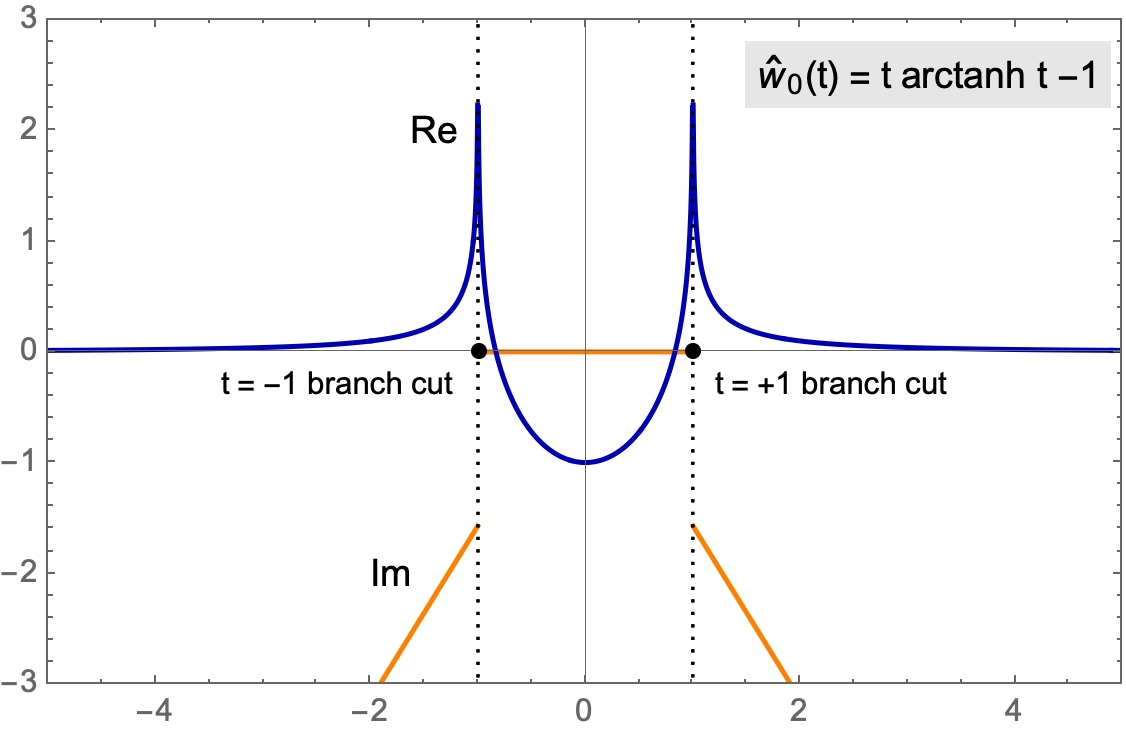}\;
\caption{\label{fig_hatw0}
Salient features of the Borel-plane representation~(\ref{eq:hattildev0}) of the universal gauge-sector background profile $y_0^G$ of Eq.~(\ref{eq:seedsintro}) in the ``ideal'' parametrization of Sec.~\ref{sec:fourthsubstitution}.} 
\end{figure}

Concerning the Borel-plane/resurgence approach to the formalism that, in  Secs.~\ref{sec:MNBPS} and~\ref{sec:NBPS}, we have been entertaining solely in the Laplace plane, there is actually a yet better way of parametrising $y$ and $z$ than any of the three different options discussed so far (in   Secs.~\ref{sec:asymptoticbackground}, \ref{sec:fRgRapproach} and \ref{sec:offsetsinboth}), namely
\begin{align}
y(x)&=1+x\, w(x)\,,\\
z(x)&=1+s(x)\,,
\end{align}
which may also be viewed as a variant of the background rearrangement/resummation approach  developed in~\cite{Malinsky:2026eux}, with
\be\label{eq:tildevfRrelation}
w(x)=e^{-x}f^R(x)\,.
\ee 
While this scheme is not optimal for the Laplace-plane type of a study like the one of Secs.~\ref{sec:MNBPS}-\ref{sec:NBPS} (the reason being the less transparent structure of the boundary conditions~(\ref{eq:boundaryconditionsy}) translated into those for $w_n$), it is actually very attractive from the Borel plane perspective as it shares most of the beautiful features encountered in the $(f^R,\, g^R)$-based approach of Sec.~\ref{sec:fRgRapproach}, such as integration-variable independence of the kernels of Volterra equations, providing a simplified route to their formal solutions. On top of that, it brings two more benefits: i) No exponentials appear explicitly on the RHS's of the relevant ODE's equivalent to Eqs.~(\ref{eq:gaugeprofile}) and~(\ref{eq:scalarprofile}), namely
\begin{align}
\label{eq:4thsubstitutionv}
L_4^1[w]&=
\frac{1}{x}+w^2\left(w+\frac{3}{x}\right)+s(2+s)\left(w+\frac{1}{x}\right)\,,\\
\label{eq:4thsubstitutions}
L_4^{2\beta}[s]&=
\frac{2}{x^2}
+\beta s^2(s+3)
+(1+s)w\left(2w+\frac{1}{x}\right)\,;
\end{align}
ii) The two sectors share {\em the same structure of the linear operator} on their LHS's:
\be
L_4^\alpha[f]\equiv 
f''+\frac{2}{x}f'-\left(\alpha+\frac{2}{x^2}\right)f\,,
\ee
with just different values of the real parameter $\alpha$ therein.
This, for the MNBPS monopole, provides perhaps the most convenient starting point of the Borel-plane triangular iteration  
\begin{align}
\label{eq:4thsubstitutionvMNBPS}
&L_4^1[w_0]=
\frac{1}{x}\,,\quad
L_4^1[w_1]=
w_0^3+\frac{3}{x}w_0^2\,,\quad\text{etc.}
\end{align} 
Indeed, thus defined $w_0$, 
corresponding in the Borel plane to the structure\footnote{This should not be surprising because the only difference between $w$ and $f^R$ is the $e^{-x}$ factor, cf.~(\ref{eq:tildevfRrelation}), inflicting a simple shift of formula~(\ref{eq:hatfRsolution}) from the resurgent germ at $t=-1$ to the regular germ at 0.}
\be\label{eq:hattildev0}
\hat{w}_0(t)=t\,{\rm arctanh}\,t-1\,,
\ee
features a symmetric pair of singularities at $t=\pm 1$ (see Fig.~\ref{fig_hatw0}) proliferating to other Borel-plane real-axis locations only due to convolutions of $\hat{w}_0$ with itself. 

It may be surprising to see a square of $w_0$ in Eq.~(\ref{eq:4thsubstitutionvMNBPS}) which, na\"\i vely, suggests that besides the expected singularities generated at relative distances of 2 (cf. Sec.~\ref{sec:MNBPSBorelPlane}), i.e., at odd-integers due to the tripple convolution of $\hat{w}_0$ in the corresponding Volterra equation 
\begin{align}\label{eq:Volterraforhattildev}
(t^2-1)\hat{w}_1(t)-2t \int_0^t {\rm d}s\, \hat{w}_1(s)&=[\hat{w}_0*\hat{w}_0*\hat{w}_0](t)\\
&+3[\hat{w}_0*\hat{w}_0*1](t)\,,\nn
\end{align}
additional singularities might also emerge at even integers from the double convolution therein (thus potentially ruining the equidistant $\Delta t_{\rm sing.}=2$ picture developed in~\cite{Malinsky:2026eux}). This, however, is not really the case due to the leading $-1$ in $\hat{w}_0$, cf.~(\ref{eq:hattildev0}); denoting $c(t)\equiv t\,{\rm arctanh}\,t$, the two non-linear terms in Eq.~(\ref{eq:Volterraforhattildev}) cancel to a large degree to produce only
\be  
2[1*1*1](t)+[c*c*c](t)\,,  
\ee 
which, at even integers, is perfectly regular. A similar cancellation is  expected also at higher orders. To this end, notice how much simpler is the structure of
\be
c(t)=\tfrac{1}{2}t\left[\log(1+t)-\log(1-t)\right]
\ee 
compared to that of ${}_2F_1\left(e^{-i\pi/3},e^{i\pi/3};1;-\tfrac{t}{2}\right)$, the Borel-plane structure underpinning the gauge-sector asymptotic background of the expansion developed in Sec.~\ref{sec:asymptoticbackground} that we have started with.  

\section{Conclusions}\label{sec:conclusions}
In this work, we have explored in great detail the intriguing resurgent structure of the classical `t~Hoof-Polyakov monopole profiles reported recently in letter~\cite{Malinsky:2026eux}. The known local transseries expansions of their profile functions, driven by a highly singular system of ordinary differential equations, have been complemented by a uniformly convergent global functional perturbative scheme based on a set of non-trivial background configurations obeying the desired boundary conditions at both limits of their physical domain simultaneously. 

Their remarkably simple analytic form, first identified in~\cite{Malinsky:2026eux} via partial resummation of an asymptotic expansion based on a ``na\"\i ve'' hypergeometric Borel-plane background, has been carefully rederived in both gauge and scalar sectors, and verified by an explicit Laplace-plane calculation.
The fast uniform convergence of thus constructed series of approximations to the ``exact'' numerical solutions has been illustrated at several examples, including the maximally non-BPS monopole corresponding to the $\beta\to\infty$ limit. For this setting, beautiful analytic prescriptions for the a-priori unknown parameters $A_\infty$ and $B_\infty$ governing the local expansions of its profile at infinite and zero radii have been obtained and used for an explicit demonstration of consistency of the scheme. 

\section*{Acknowledgments}
The work has been performed with the support from the 
Charles University Research Center of Excellence 
UNCE/24/SCI/016 grant and from the FORTE project CZ.02.01.01/00/22\_008/0004632 co-funded by the EU and the Ministry of Education, Youth and Sports of the Czech Republic. 

\appendix
\section{MNBPS profile - asymptotic expansion\label{app:MNBPSasymptoticprofile}}
\label{app:MNBPStransseriesInfinity}
Concerning the large-$x$ expansion of $y$ obeying in the MNBPS ($\beta\to\infty$) limit Eq.~(\ref{eq:MNBPSprofile}), the first thing that comes to mind is a transseries ansatz
\begin{equation}\label{eq:InfinityTransseries}
y(x)= \sum_{m=0}^\infty \sum_{n=1}^\infty   c_{m,n}x^{-m}{e^{-n x}}\,,
\end{equation}
where the negative exponentials are motivated by the expected behaviour of Eq.~(\ref{eq:MNBPSprofile}) in the asymptotic region where it assumes a purely exponential form $y''\approx y$.
The $c_{m,n}$ coefficients are then determined from the comparison of the LHS and RHS of the full Eq.~(\ref{eq:MNBPSprofile}), namely
\begin{align}
c_{m,n}(1-n^2)&=2n(m-1)c_{m-1,n}\\
&+\left[(m-1)(m-2)+1\right]c_{m-2,n}-S^c_{m,n}\nn
\end{align}
where
\be
S^c_{m,n}\equiv\sum_{\substack{k_1+k_2+k_3=m-2;\; k_i\geq 0 \\ l_1+l_2+l_3=n;\;l_i > 0}}c_{k_1,l_1}c_{k_2,l_2}c_{k_3,l_3}\,,
\ee
stemming from the $y^3/x^2$ non-linearity therein, is a sum over indices corresponding to all possible non-negative partitions of $m-2$ and all possible positive partitions of $n$. One can check easily that $c_{m,n}$ can be non-zero only for odd $n\geq 1$, $m\geq n-1$ and that the value of the ``seed element'' $C\equiv c_{0,1}$ remains undetermined.

Interestingly, due to $S_{m,1}=0$, the $c_{m,1}$ sequence can be written in a closed form 
\be\label{eq:definitioncm1}
c_{m,1}=C \left(-\frac{1}{2}\right)^m\frac{1}{m!}\prod_{k=1}^m (k^2-k+1)\,.
\ee 
The bracket under the product sign can be recast like $k^2-k+1=(k-e^{i\pi/3})(k-e^{-i\pi/3})$ and, hence, the product in~(\ref{eq:definitioncm1}) is nothing but a product of two Pochhammer symbols  $(p)_k\equiv p(p+1)(p+2)\ldots (p+k-1)$,
\begin{align}
\prod_{k=1}^m (k-e^{i\pi/3})(k+e^{i\pi/3})=(1-e^{i\pi/3})_m (1-e^{-i\pi/3})_m\,.\nn
\end{align}
Note also that $e^{\pm i\pi/3}$ are rather special numbers as they obey
\begin{align}
(1-e^{i\pi/3})_m (1-e^{-i\pi/3})_m=(e^{-i\pi/3})_m (e^{i\pi/3})_m\,.\nn
\end{align}
This is very interesting because of the close similarity of thus rewritten $c_{m,1}$ coefficients 
\begin{align}\label{eq:cm1intermsofPochhammer}
c_{m,1}=C \left(-\frac{1}{2}\right)^m\frac{1}{m!}(e^{-i\pi/3})_m (e^{i\pi/3})_m
\end{align}
with the general coefficients of the local expansion of the ${}_2F_1$ Gaussian hypergeometric function at the origin, namely
\be\label{eq:general2F1}
{}_2F_1(\alpha,\beta;\gamma;z)=\sum_{m=0}^\infty \frac{1}{m!}\frac{(\alpha)_m (\beta)_m}{(\gamma)_m}z^m\,.
\ee 
Matching~(\ref{eq:cm1intermsofPochhammer}) to~(\ref{eq:general2F1}) and recalling $(1)_m=m!$, one arrives at an intriguing relation
\be\label{eq:hypergeometricsindm}
{}_2F_1(e^{-i\pi/3},e^{i\pi/3};1;-\tfrac{z}{2})=\sum_{m=0}^\infty d_m z^m\,,
\ee
where
\be\label{eq:dmcoefficients}
d_m=\frac{1}{C}\frac{c_{m,1}}{m!}\,.
\ee

\subsection*{Borel resummation of the $n=0$ tower}
\begin{figure}[t]
  \includegraphics[width=0.42\textwidth]{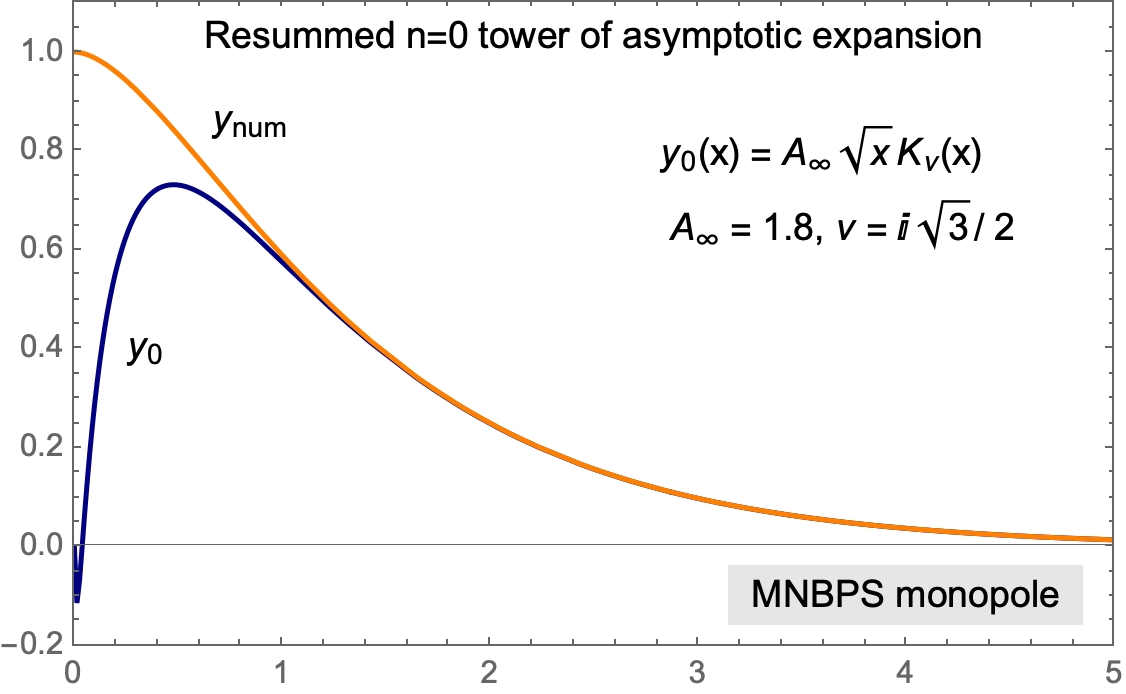}\;
\caption{\label{fig:y0asymptoticprofile}
Comparison of the resummed leading asymptotic (gauge) profile of the MNBPS monopole $y_0$ of Eq.~(\ref{eq:MNBPSasymptoticprofile}) (in blue) to the  ``exact'' numerical solution $y_{\rm num}$ of Eq.~(\ref{eq:MNBPSprofile}) (in orange) for the value of the free parameter $A_\infty=1.8$. As expected, $y_0$ exhibits a perfect match to $y_{\rm num}$ for large $x$ by construction, see also~\cite{Dunne:2026hfx}. Its behaviour around the origin is relatively wild though, see also Sec.~\ref{sec:inadequacy}.} 
\end{figure}

This all means that the expansion~(\ref{eq:InfinityTransseries}) we have started with can be eventually written as 
\begin{equation}\label{eq:InfinityTransseriesv2}
y(x)= \sum_{m=0}^\infty \sum_{n=0}^m a_{m,n}x^{-m}{e^{-(2n+1)x}}\,,
\end{equation}
where 
\be\label{eq:accorrespondence}
a_{m,n}\equiv c_{m,2n+1}\,,
\ee
with the coefficients of its lowest-exponential ($n=0$) tower given by $a_{m,0}=C d_m m!$. Hence, the Gauss hypergeometric function ${}_2F_1(e^{-i\pi/3},e^{i\pi/3};1;-\tfrac{z}{2})$ is essentially the Borel resum of the inverse-power ``core'' of the $n=0$ tower of $y$ (to be denoted $y_0$), written conveniently as
\be\label{eq:y0intermsoff0}
y_0(x)=x e^{-x} f_0(x)
\,,
\ee
with 
\be
f_0(x)\equiv\sum_{m=0}^\infty \frac{a_{m,0}}{x^{m+1}}=C \sum_{m=0}^\infty m! \frac{d_{m}}{x^{m+1}}\,.
\ee
In other words, the $n=0$ tower of the expansion~(\ref{eq:InfinityTransseriesv2}) can be resummed as a Laplace transform of the ${}_2F_1$ function (which, in the Borel language is nothing but the analytic continuation of 
\be\label{eq:f0resummation}
\hat{f}_0(t)\equiv C \sum_{m=0}^\infty d_m t^m
\ee 
to the entire complex plane),
\begin{equation}
y_0(x)=A_\infty\sqrt{\tfrac{\pi}{2}}\, x\, e^{-x}\int_0^\infty\!\! e^{-x t}{}_2F_1(e^{-i\pi/3},e^{i\pi/3};1;-\tfrac{t}{2})\,,
\end{equation}
with $A_\infty\equiv C\sqrt{\frac{2}{\pi}}$ which, by virtue of the general identity connecting its RHS to the modified Bessel $K$-functions of order $\nu=i\sqrt{3}/2$, cf.~\cite{Dunne:2025mye},
\begin{equation}
\sqrt{x}K_\nu(x)=\sqrt{\tfrac{\pi}{2}}\,x e^{-x}\int_0^\infty\!\! e^{-x t}{}_2F_1(e^{-i\pi/3},e^{i\pi/3};1;-\tfrac{t}{2})\,,
\end{equation}
provides a simple resummed formula   
\begin{equation}\label{eq:MNBPSasymptoticprofile}
y_0(x)=A_\infty\sqrt{x}K_{i\sqrt{3}/2}(x)\,.
\end{equation}
This was one of the starting points of the resurgent approach formulated in letter~\cite{Malinsky:2026eux}.
A comparison of  $y_0$ of Eq.~(\ref{eq:MNBPSasymptoticprofile}) to the ``exact'' numerical solution of Eq.~(\ref{eq:MNBPSprofile}) is given in Fig.~\ref{fig:y0asymptoticprofile}.

For completeness, let us write down first few coefficients of the $n=0,1,2$ towers of the expansion~(\ref{eq:InfinityTransseriesv2}) to illustrate its clear asymptotic nature:
\begin{align}
\label{eq:amnnumericaltowers}
&\frac{a_{m,0}}{C}=\left\{1,-\tfrac{1}{2},\tfrac{3}{8},-\tfrac{7}{16},\tfrac{91}{128},-\tfrac{1911}{1280},\tfrac{19747}{5120},-\tfrac{121303}{10240},\ldots\right\},\nn\\
&\frac{a_{m+2,1}}{C^3}=\left\{\tfrac{1}{8},-\tfrac{3}{8}, \tfrac{31}{32}, -\tfrac{335}{128}, \tfrac{7977}{1024}, -\tfrac{33191}{1280},\tfrac{247491}{2560},\ldots\right\},\nn\\
&\frac{a_{m+4,2}}{C^5}=\left\{\tfrac{1}{64},-\tfrac{17}{192}, \tfrac{1633}{4608}, -\tfrac{5927}{4608}, \tfrac{1023737}{221184},\ldots\right\}.
\end{align}
It is perhaps also worth mentioning that there is a shortcut to the leading resummed profile~(\ref{eq:MNBPSasymptoticprofile}) exploiting the observation that, upon $y=\sqrt{x}a(x)$, one does not need to drop the entire $x^{-2}$-proportional structure on the RHS of the MNBPS profile ODE~(\ref{eq:MNBPSprofile}), just the $y^3/x^2$ term, and still retain exact solvability of thus constructed ODE for $a(x)$, which turns out to be of the modified Bessel type, see also formula~(\ref{BesselODE}).  
 
\section{MNBPS profile - local expansion at 0\label{app:MNBPSzeroprofile}}
Interestingly, even the expansion of the MNBPS gauge profile around $x=0$ requires a non-trivial (transseries) ansatz. Indeed, attempting a pure Taylor expansion  
\be
y(x)=\sum_{m=0}^\infty b_m x^m\,,
\ee 
the constraints on the $b_m$ coefficients enforced by Eq.~(\ref{eq:MNBPSprofile}) become singular (no solution would exist) already at the $x^0$ level. This is due to the infamous Frobenius resonance among different sectors which, in turn, requires inclusion of logarithms. Doing that, one eventually arrives at
\begin{equation}\label{eq:yexpansionaround0}
y(x)=\sum_{m=0}^\infty \sum_{n=0}^m b_{m,n}x^{2m}(\log x)^n\,,
\end{equation}
where 
the comparison of the LHS and RHS of Eq.~(\ref{eq:MNBPSprofile}) yields
\begin{align}\label{eq:bmnrecursion}
b_{m,n}&=S^b_{m,n}+\left[(2m+2)(2m+1)+1\right]b_{m+1,n}\\
&+(n\!+\!1)(4m\!+\!3)b_{m+1,n+1}+(n\!+\!2)(n\!+\!1)b_{m+1,n+2}\nn
\end{align}
where
\be
S^b_{m,n}=\sum_{\substack{k_1+k_2+k_3=m+1;\; k_i\geq 0 \\ l_1+l_2+l_3=n;\;l_i \geq 0,\\ l_1<i_1,\, l_2<i_2,\, l_3<i_3}}b_{k_1,l_1}b_{k_2,l_2}b_{k_3,l_3}\,,
\ee
again stems from the $y^3/x^2$ nonlinearity therein. Setting $b_{0,0}=1$ from the boundary condition at $x=0$, $b_{1,0}$ remains undetermined and plays the role of the expected a-priori unknown real parameter (to be denoted by $B_\infty$) which is nothing but the $x=0$ counterpart of the free parameter $A_\infty$ of the $x\to \infty$ expansion of Appendix~\ref{app:MNBPSasymptoticprofile}; all ``higher'' $b_{m,n}$ then follow from the recursion~(\ref{eq:bmnrecursion}).

\begin{figure}[t]
  \includegraphics[width=0.42\textwidth]{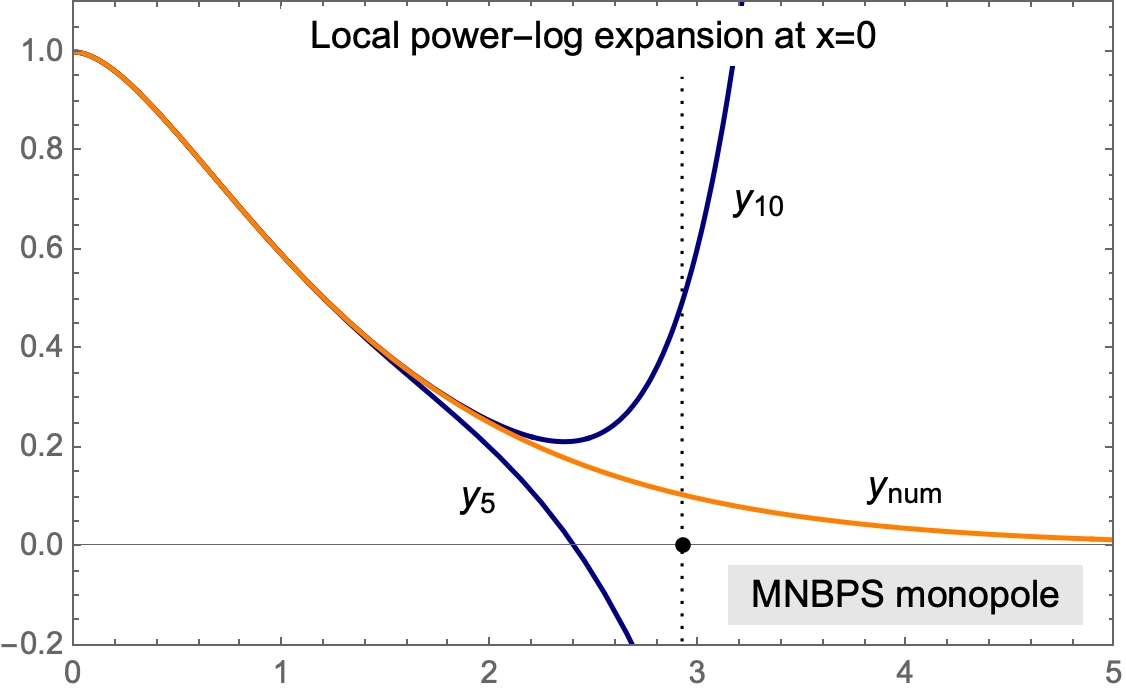}\;
\caption{\label{fig:y0frobeniusexpansion}
Numerical behaviour of the local power-log expansion~(\ref{eq:yexpansionaround0}), cf. also~(\ref{eq:Frobeniusat0intro}), with $B_\infty=-0.484$ for two different truncations corresponding to $m_{\rm max}=5$ and $m_{\rm max}=10$ (in blue), respectively, compared to the numerical solution $y_{\rm num}$ (in orange). The vertical dashed line corresponds roughly to the convergence radius of~(\ref{eq:yexpansionaround0}), see Appendix~\ref{app:MNBPSzeroprofile}.} 
\end{figure}

For illustration, the first few $b_{m,n}$ coefficients read
\begin{align}
&b_{0,0}=1\,,\\
&b_{1,0}=B_\infty\,,\;b_{1,1}=\tfrac{1}{3}\,,
\nn\\
&b_{2,0}=\tfrac{3}{10} B_\infty^2-\tfrac{1}{25} B_\infty+\tfrac{1}{375}\,,\,b_{2,1}=\tfrac{1}{5}B_\infty-\tfrac{1}{75}\,,\,b_{2,2}=\tfrac{1}{30}\,,
\nn\\
&b_{3,0}=\tfrac{1}{10} B_\infty^3-\tfrac{13}{350} B_\infty^2+\tfrac{159}{24500}B_\infty-\tfrac{2839}{6174000}\,,\nn\\
&b_{3,1}=\tfrac{1}{10} B_\infty^2-\tfrac{13}{525} B_\infty+\tfrac{53}{24500}\,,\nn\\
&b_{3,2}=\tfrac{1}{30}B_\infty-\tfrac{13}{3150}\,,\;
b_{3,3}=\tfrac{1}{270}\,,\nn
\end{align}
and so on.
The numerical behaviour of a truncated version of expansion~(\ref{eq:yexpansionaround0}) for a pair of maximum $m$'s, namely, $m_{\rm max}=5$ and $m_{\rm max}=10$, is depicted in Fig.~\ref{fig:y0frobeniusexpansion}. Note that the $B_\infty$ parameter,  fitted to be
\be\label{eq:Binftyfitted}
B_\infty\approx -0.484\,,
\ee
obeys a closed-form relation~(\ref{eq:Bfull}), cf. Sec.~\ref{sec:Binfinityexact}.  
\subsection*{Approximate upper limit on the convergence radius}  
As it is to be expected, the local Frobenius expansion~(\ref{eq:yexpansionaround0}) has got only a finite radius of convergence. Barring large all-orders cancellations (i.e., a structural conspiracy) among its individual terms, it can be roughly estimated from the progression of its simplest (and $B_\infty$-independent) diagonal  coefficients $b_{m,m}$. By virtue of Eq.~(\ref{eq:bmnrecursion}), these form a sector which is recursively closed:
\be
\left[2m(2m-1)+1\right]b_{m,m}\;=\!\!\!\!\!\!\sum_{\substack{i+j+k=m\\ i\geq 0,\,j\geq 0,\,k\geq 0}}b_{i,i}b_{j,j}b_{k,k}\,.
\ee
The solution of this relation for $b_{0,0}=1$ can be shown to be of a geometric nature, namely, $b_{m,m}=\sqrt{8}\,q^m$ with the quotient determined numerically to be $q\approx 0.109374$. Hence, the large-$m$ terms of the highest-log sub-series of~(\ref{eq:yexpansionaround0}) behave as $\sqrt{8}(q\, x^2 \log x)^m$ 
and, hence, the radius of convergence of this sub-sector in isolation is given by 
\be
|x^2 \log x| < q^{-1}\,,
\ee
which can be solved in terms of the Lambert function $W$ as 
$|x|<\exp\left[\tfrac{1}{2}W\left(\tfrac{2}{q}\right)\right]\approx 2.921$.      

\section{The fundamental ODE - vector sector}\label{app:fundamentalODE}
In this Appendix we present two complementary analytic accounts of the solutions of the fundamental ODE 
\begin{equation}\label{eq:fundamentalODE}
{
v''(x)-\left(1+\frac{2}{x^2}\right)v(x)=r(x)
}\,.
\end{equation}
governing all levels of the general vector-profile expansions (of both the MNBPS as well as the general NBPS cases) developed in Secs.~\ref{eq:generalvsexpansion}, see Eqs.~(\ref{eq:MNBPSvectorprofileinv}) and (\ref{eq:ODEforsn}), first from the perspective of the Laplace plane and then from the resurgence/Borel-plane viewpoint. The beauty of the specific parametrisation~(\ref{eq:vsparametrization}) (corresponding to the partial resummation of the expansion based on the asymptotic background~$f_0$ of Eq.~(\ref{eq:Laplacef0asymptotics}), cf.~\cite{Malinsky:2026eux}) is the simplicity of the relevant Borel-plane Volterra equation that, as we shall see, turns out to be equivalent to a 1st order ODE easily solvable in quadratures (and even analytically for the $n=0$ basis of the expansion tower corresponding to $r_0(x)=1$, cf. Eq.~(\ref{eq:r0})). 
   
\subsection{Laplace-plane viewpoint}
The general solution of Eq.~(\ref{eq:fundamentalODE}) can be written as 
\be 
{
v(x)=h(x)+p(x)}\,,
\ee
where $h$ is the solution to its homogeneous version 
\begin{equation}\label{eq:homogeneousfundamentalODE}
h''(x)-\left(1+\frac{2}{x^2}\right)h(x)=0\,,
\end{equation}
and $p$ is the particular solution associated to $r(x)$, obtained by the variation of constants method.

\subsubsection{Homogeneous solution\label{app:homogeneoussolutionvector}}
\begin{figure}[t]
  \includegraphics[width=0.42\textwidth]{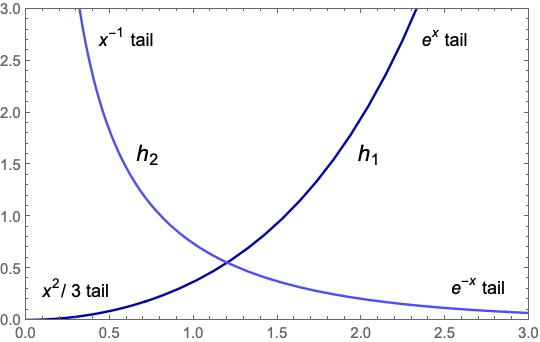}\;
\caption{\label{fig:h1h2}
The two fundamental modes~(\ref{eq:h1vector})-(\ref{eq:h2vector}) of Eq.~(\ref{eq:fundamentalODE}), closely related to modified Bessel functions of order $3/2$, see Appendix~\ref{app:homogeneoussolutionvector}, that are the core objects of the global expansion method contemplated in this work.} 
\end{figure}

As for the homogeneous part, employing $h(x)\equiv\sqrt{x}\,a(x)$, Eq.~(\ref{eq:homogeneousfundamentalODE}) is readily transformed into an ODE for modified Bessel functions
\begin{equation}\label{BesselODE}
x^2 a''(x)+x a'(x)-(x^2+\nu^2)a(x)=0\,,
\end{equation}
with $\nu^2=9/4$; hence, the ``canonical pair'' of fundamental solutions of Eq.~(\ref{eq:homogeneousfundamentalODE}) is
\begin{align}
\label{eq:fundB1}
\sqrt{x}I_{3/2}(x)&=\sqrt{\frac{2}{\pi}}\left(-\frac{\sinh x}{x}+\cosh x\right)\,,
\\
\label{eq:fundB2}
\sqrt{x}K_{3/2}(x)&=\sqrt{\frac{\pi}{2}}\left(\frac{e^{-x}}{x}+e^{-x}\right)
\,,
\end{align}
where $I_{3/2}$ and $K_{3/2}$ are the modified Bessel functions of a real half-integer order that, as indicated, can be represented in terms of elementary functions!
The solution of Eq.~(\ref{eq:homogeneousfundamentalODE}) can thus be written as
\be\label{eq:homogeneoussolution}
{
h(x)=C_1 h_1(x)+C_2 h_2(x)
}\,,
\ee
which utilizes a yet more convenient (but entirely equivalent) form of the pair of fundamental solutions~(\ref{eq:fundB1})-(\ref{eq:fundB2}), 
\begin{align}
\label{eq:h1vector}
h_1(x)&\equiv-\frac{\sinh x}{x}+\cosh x\,, \\
\label{eq:h2vector}
h_2(x)&\equiv\frac{e^{-x}}{x}+e^{-x}\,,
\end{align}
see also~\cite{Bais:1976}.
For illustration, the graphs of these functions are depicted in Fig.~\ref{fig:h1h2}.

Note, in particular, that due to the reciprocity of the overall numerical factors in (\ref{eq:fundB1}) and (\ref{eq:fundB2}) the Wronskians of both fundamental pairs (\ref{eq:fundB1})-(\ref{eq:fundB2}) and (\ref{eq:h1vector})-(\ref{eq:h2vector}), that should be constant due to the absence of a first derivative term from Eq.~(\ref{eq:homogeneousfundamentalODE}), are indeed the same,
\begin{align}
W\left[\sqrt{x}K_{3/2}(x),\sqrt{x}I_{3/2}(x)\right]=W\left[h_1,h_2\right](x)=-1\,.
\nn
\end{align}
\subsubsection{Particular solution}
This eventually yields a very simple form of the particular solution of Eq.~(\ref{eq:fundamentalODE}), namely
\be\label{eq:particularsolutionvector}
p(x)=h_1(x)\int_{x_0}^x h_2(z)r(z){\rm d}z-h_2(x)\int_{x_0}^x h_1(z)r(z){\rm d}z
\,,
\ee
where the specific choice of $x_0$ is irrelevant from the full solution perspective as any change in it can always be absorbed into a redefinition of the $C_1$ and $C_2$ constants in the homogeneous part. 
\subsection{Borel-plane/resurgence viewpoint}
In the parametrization of Ref.~\cite{Malinsky:2026eux} corresponding to the resummed vector profile $y(x)=1+x e^{-x} f^R(x)$ therein, one has 
\be\label{eq:keyreparametrization}
v(x)=x e^{-x} f^R(x)
\ee 
and, thus, the Borel-plane equivalent of the ODE~(\ref{eq:fundamentalODE}) written in terms of $\hat{f}^R$ reads
\be\label{eq:resummedvectorVolterra}
t(t+2)\hat{f}^R(t)-2(t+1)\int_0^t {\rm d}s\,\hat{f}^R(s)=\hat{r}^f(t)\,,
\ee
with $\hat{r}^f(t)$ denoting the Borel-plane representation of 
$\frac{e^x}{x}r(x)$ where $r(x)$ is the RHS of Eq.~(\ref{eq:fundamentalODE}).

Two things are worth noting here: i) The non-trivial kernel $K(t,s)=t-3s-2$ of Eq.~(7) in Ref.~\cite{Malinsky:2026eux}, corresponding to the original $\hat{f}$, becomes an $s$-independent structure $-2(t+1)$ in the resummed scheme that one could take out of the Volterra integral; and ii) Eq.~(\ref{eq:resummedvectorVolterra}) reveals two singularities in the Borel plane; thus, two germs (at $t=0$ and $t=-2$) are developed. In terms of 
\be
F(t)\equiv\int_0^t {\rm d}s\,\hat{f}^R(s)\,,
\ee
Eq.~(\ref{eq:resummedvectorVolterra}) becomes equivalent to a 1st order linear ODE
\be\label{eq:resummedvectorVolterraODE}
t(t+2)F'(t)-2(t+1)F(t)=\hat{r}^f(t)\,.
\ee
Its general solution can be again obtained by looking first at the homogeneous case which is separable and trivially integrated into 
\be
F_h(t)=C t(t+2)\quad \text{and, hence}\quad \hat{f}_h^R(t)=2C (t+1)\,,
\ee
while the particular solution obtained by the variation of $C$ 
reads 
\be
F_p(t)=t(t+2)\int_{t_0}^t{\rm d}s\frac{\hat{r}^f(t)}{s^2(s+2)^2}\,,
\ee 
with $t_0$ serving as an integration constant. 
This, in turn yields for $\hat{f}^R(t)=F_h'(t)+F_p'(t)$:   
\be\label{eq:generalhatfR}
\hat{f}^R(t)=2C(t+1)+\frac{\hat{r}^f(t)}{t(t+2)}+2(t+1)\int_{t_0}^t{\rm d}s\frac{\hat{r}^f(t)}{s^2(s+2)^2}\,.
\ee 
Two comments may be worth here: i) There is only one integration constant in~(\ref{eq:generalhatfR}) because $C$ and $t_0$ sum to only one constant structure in front of $(t+1)$. ii) The fact that the ODE~(\ref{eq:resummedvectorVolterraODE}) is only 1st order (and, thus, ``optically'' possesses only one fundamental mode), while the two ODE's~(\ref{eq:h1vector})-(\ref{eq:h2vector}) of the equivalent Laplace-plane problem possess two independent fundamental modes, is not a contradiction. The point is that the expression~(\ref{eq:generalhatfR}) defines local data of {\em two} germs identified above and, in fact, the Laplace transforms of $\hat{f}_h^R$ expressed in their local coordinates $t=-2+u_1$ and $t=u_2$ define two independent structures
\begin{align}
f^R_1(x)&={\cal L}_{(-2)}[-1+u_1](x)=e^{2x}\left(-\frac{1}{x}+\frac{1}{x^2}\right)\,,\\
f^R_2(x)&={\cal L}_{(0)}[1+u_2](x)=\frac{1}{x}+\frac{1}{x^2}\,,
\end{align}
that, multiplied by $x e^{-x}$, cf.~(\ref{eq:keyreparametrization}), provide a fundamental system equivalent to~(\ref{eq:h1vector}) and~(\ref{eq:h2vector}).              

\section{The fundamental ODE - scalar sector\label{app:fundamentalODE'scalar}}
In this Appendix we provide the same analysis as that of Appendix~\ref{app:fundamentalODE}, but for the fundamental ODE\footnote{In order to keep the notation simple, we use here the same generic symbols $h$, $p$ and $r$ for the relevant quantities as we did in Appendix~\ref{app:fundamentalODE}; in other words, the scope of these symbols is restricted to the context of the corresponding Appendix only.}
\begin{equation}\label{eq:fundamentalODE'scalar}
{
s''(x)+\frac{2 s'(x)}{x}-\left(2\beta+\frac{2}{x^2}\right)s(x)=r(x)
}
\end{equation}
governing the scalar sector perturbative expansion~(\ref{eq:ODEforsn}). 
\subsection{Laplace-plane viewpoint}
As before, we write the general solution to Eq.~(\ref{eq:fundamentalODE'scalar}) in terms of the homogeneous and particular solutions,
\be
s(x)=h(x)+p(x)\,,
\ee 
where $h$ is the general solution of 
\begin{equation}\label{eq:fundamentalODE'scalarhomogeneous}
{
s''(x)+\frac{2 s'(x)}{x}-\left(2\beta+\frac{2}{x^2}\right)s(x)=0\,,
}
\end{equation}
and $p$ associated to a non-zero RHS of Eq.~(\ref{eq:fundamentalODE'scalar}) is obtained as usual by variation of its constants.

\subsubsection{Homogeneous solution}
Here the trick that transforms Eq.~(\ref{eq:fundamentalODE'scalarhomogeneous}) into a standard form, for which an analytic solution is known, is $h(x)\equiv\,a(\sqrt{2\beta} x)/\sqrt{x}$. Remarkably enough, even in this case $a$  conforms the ODE~(\ref{BesselODE}) for the modified Bessel functions of order $3/2$, and, hence, the pair of fundamental solutions of~(\ref{eq:fundamentalODE'scalarhomogeneous}) can be again written in terms of the $I_{3/2}$ and $K_{3/2}$ functions, or even better in terms of their equivalents~(\ref{eq:h1vector})-(\ref{eq:h2vector}), namely  
\begin{align}
\label{eq:h1scalar}
\frac{h_1(\sqrt{2\beta}x)}{x}&=-\frac{\sinh \sqrt{2\beta}x}{\sqrt{2\beta}x^2}+\frac{\cosh \sqrt{2\beta}x}{x}\,, \\
\label{eq:h2scalar}
\frac{h_2(\sqrt{2\beta}x)}{x}&=\frac{e^{-\sqrt{2\beta}x}}{\sqrt{2\beta}x^2}+\frac{e^{-\sqrt{2\beta}x}}{x}\,.
\end{align}
Note that, unlike in the vector case of Appendix~\ref{app:fundamentalODE}, their Wronskian 
\begin{align}
W\left[\frac{h_1(\sqrt{2\beta}x)}{x},\frac{h_2(\sqrt{2\beta}x)}{x}\right](x)=-\frac{\sqrt{2\beta}}{x^2}
\end{align}
is non-trivial and, as such, it will appear explicitly in the particular solution prescription.
   
\subsubsection{Particular solution}
Given this, the shape of the particular solution in the scalar sector is only slightly more complicated than that for the vector profile~(\ref{eq:particularsolutionvector}), namely, 
\begin{align}
p(x)&=\frac{h_1(\sqrt{2\beta}x)}{x}\int_{x_0}^x {\rm d}z\frac{h_2(\sqrt{2\beta}z)}{z\, W(z)}r(z)\\
&-\frac{h_2(\sqrt{2\beta}x)}{x}\int_{x_0}^x {\rm d}z\frac{h_1(\sqrt{2\beta}z)}{z\, W(z)}r(z)\nn
\,.
\end{align}

\subsection{Borel-plane/resurgence viewpoint}
The Volterra equation equivalent to Eq.~(\ref{eq:fundamentalODE'scalar}) reads 
\be\label{eq:resummedscalarVolterra}
(t^2-2\beta)\hat{s}(t)-2t\int_0^t {\rm d}z\hat{s}(z)=\hat{r}^s(t)\,,
\ee
where $\hat{r}^s(t)$ is the Borel-plane equivalent of $r(x)$ therein. Note that this Equation has two singularities at $t=\pm\sqrt{2\beta}$. Again, the two integrals corresponding to the $2 s'(x)/x$ and $-2 s(x)/x^2$ structures in Eq.~(\ref{eq:fundamentalODE'scalar}) combine in such a way to produce a Volterra kernel $-2t$ that does not depend on the integration variable. Thus, Eq.~(\ref{eq:resummedscalarVolterra}) is equivalent to a 1st order ODE  
\be
(t^2-2\beta)S'(t)-2tS(t)=\hat{r}(t)
\ee
for
\be
S(t)=\int_0^t {\rm d}z\hat{s}(z)\,.
\ee
Its homogeneous version is again separable and easy to integrate
\be
S_h(t)=C(t^2-2\beta)\,,
\ee
and the particular solution obtained by variation of $C$ reads 
\be
S_p(t)=(t^2-2\beta)\int_{t_0}^t{\rm d}s\frac{s\, \hat{r}(s)}{(s^2-2\beta)^2}\,.
\ee 
Note that as in the vector case the full $S(t)$ also depends only on one combination of $C$ and $t_0$. These findings are readily translated into 
\be
\hat{s}(t)=2C t+\frac{t\, \hat{r}(t)}{(t^2-2\beta)}+2t\int_{t_0}^t{\rm d}s\frac{s\, \hat{r}(s)}{(s^2-2\beta)^2}\,.
\ee
The two fundamental modes equivalent to~(\ref{eq:h1scalar})-(\ref{eq:h2scalar}) are again obtained from the single homogeneous solution $\hat{s}_h(t)=2Ct$ defining local data of the two germs at $t=\pm\sqrt{2\beta}$. The Laplace transforms of $\hat{s}_h$ expressed in their local coordinates $t=-\sqrt{2\beta}+u_1$ and $t=\sqrt{2\beta}+u_2$ define two independent structures
\begin{align}
s_1(x)&={\cal L}_{(-\sqrt{2\beta})}[u_1-\sqrt{2\beta}](x)=e^{\sqrt{2\beta}x}\left(-\frac{\sqrt{2\beta}}{x}+\frac{1}{x^2}\right)\,,\\
s_2(x)&={\cal L}_{(\sqrt{2\beta})}[u_2+\sqrt{2\beta}](x)=e^{-\sqrt{2\beta}x}\left(\frac{\sqrt{2\beta}}{x}+\frac{1}{x^2}\right)\,,
\end{align}
which, as expected, are just linear combinations of $h_1(\sqrt{2\beta}x)/x$ and $h_2(\sqrt{2\beta}x)/x$ of Eqs.~(\ref{eq:h1scalar})-(\ref{eq:h2scalar}). 
\section{Resurgent structure of the BPS profile}
The BPS monopole is the only case in which the gauge-profile asymptotics is not of the form~(\ref{eq:asymptoticbackground}). In coordinates~(\ref{eq:subst1vector}), it has a trivial geometric-series-like expansion
\begin{equation}\label{eq:BPStower}
f_{\rm BPS}(x)={2}(1-e^{-2x})^{-1}=2(1+e^{-2x}+e^{-4x}+\ldots)\,,
\end{equation}
which can be viewed as a resurgent structure emerging from a ``Dirac comb'' structure with delta-function singularities at non-negative even integers in the Borel plane. In this respect, the BPS solution somewhat resembles the MNBPS one discussed in detail in Sec.~\ref{sec:MNBPSBorelPlane}, see also~\cite{Malinsky:2026eux}. 
    
Interestingly, the $f_{\rm BPS}$ function above can also be Laurent-expanded around $x=0$, 
\begin{equation}\label{eq:fBPS}
f_{\rm BPS}(x)=\tfrac{1}{x}+1+\sum_{n=1}^\infty \frac{B_{2n}}{(2n)!}(2x)^{2n-1}\,,
\end{equation}
(with $B_{2n}$ denoting even Bernoulli numbers) which -- barring the pole term -- is locally convergent. Its Borel-plane counterpart, up to a constant from the regular germ, is
$\hat{f}_{\rm BPS}(t)=\tfrac{1}{2}\coth(t/2)-1/t$,
with singularities at $t=2\pi i n$ with $n=\pm 1, \pm 2$ etc, i.e., a $\Delta t=2$ equidistant singularity pattern emerges again (modulo a benign rotation of the integration contour). One can also notice that $\hat{f}_{\rm BPS}$ is in fact functionally identical to the Laplace-plane shape of its scalar counterpart, which is interesting.      


\end{document}